\pdfoutput=1
\documentclass[12pt,a4paper]{article}
\usepackage{ifthen} % for conditional statements
\newboolean{pdflatex}
\setboolean{pdflatex}{true} % False for eps figures 

\newboolean{articletitles}
\setboolean{articletitles}{true} % False removes titles in references

\newboolean{uprightparticles}
\setboolean{uprightparticles}{false} %True for upright particle symbols

\def\paperauthors{LHCb collaboration} % Leave as is for PAPER, CONF and FIGURE
\def\paperasciititle{Measurement of D0-D0bar mixing and search for CP violation with D0 -> K+ pi- decays} % Set ASCII title here !! MAKE sure it's only ASCII characters !! 
\def\papertitle{Measurement of \Dz--\Dzb mixing \\ and search for \CP violation \\ with \decay{\Dz}{\Kp\pim} decays} % Latex formatted title
\def\paperkeywords{{High Energy Physics}, {LHCb},{flavour physics}, {charm physics}, {CP violation}, {time-dependent CP violation}, {indirect CP violation}, {oscillation}} % Comma separated list
\def\papercopyright{\the\year\ CERN for the benefit of the LHCb collaboration} % new since 9/Apr/2018
\def\paperlicence{CC BY 4.0 licence}
\def\paperlicenceurl{https://creativecommons.org/licenses/by/4.0/}

\usepackage[top=1in, bottom=1.25in, left=1in, right=1in]{geometry}

\usepackage{microtype}
\usepackage{lineno}  % for line numbering during review
\usepackage{xspace} % To avoid problems with missing or double spaces after
\usepackage{caption} %these three command get the figure and table captions automatically small

\usepackage{graphicx}  % to include figures (can also use other packages)
\usepackage{color}
\usepackage{colortbl}
\graphicspath{{./figs/}} % Make Latex search fig subdir for figures
\usepackage{amsmath} % Adds a large collection of math symbols
\usepackage{amssymb}
\usepackage{amsfonts}
\usepackage{upgreek} % Adds in support for greek letters in roman typeset

\newcommand*\patchAmsMathEnvironmentForLineno[1]{%
\expandafter\let\csname old#1\expandafter\endcsname\csname #1\endcsname
\expandafter\let\csname oldend#1\expandafter\endcsname\csname
end#1\endcsname
 \renewenvironment{#1}%
   {\linenomath\csname old#1\endcsname}%
   {\csname oldend#1\endcsname\endlinenomath}%
}
\newcommand*\patchBothAmsMathEnvironmentsForLineno[1]{%
  \patchAmsMathEnvironmentForLineno{#1}%
  \patchAmsMathEnvironmentForLineno{#1*}%
}
\AtBeginDocument{%
\patchBothAmsMathEnvironmentsForLineno{equation}%
\patchBothAmsMathEnvironmentsForLineno{align}%
\patchBothAmsMathEnvironmentsForLineno{flalign}%
\patchBothAmsMathEnvironmentsForLineno{alignat}%
\patchBothAmsMathEnvironmentsForLineno{gather}%
\patchBothAmsMathEnvironmentsForLineno{multline}%
\patchBothAmsMathEnvironmentsForLineno{eqnarray}%
}

\usepackage{hyperxmp}
\usepackage{mathtools}
\usepackage[pdftex,
            pdfauthor={\paperauthors},
            pdftitle={\paperasciititle},
            pdfkeywords={\paperkeywords},
            pdfcopyright={Copyright (C) \papercopyright},
            pdflicenseurl={\paperlicenceurl}]{hyperref}
\usepackage[bottom,flushmargin,hang,multiple]{footmisc}

\usepackage[all]{hypcap} % Internal hyperlinks to floats.

\usepackage{xspace} 
\usepackage{upgreek}

\def\lhcb   {\mbox{LHCb}\xspace}

\def\lhc    {\mbox{LHC}\xspace}

\def\rich   {RICH\xspace}

\def\MagUp {\mbox{\em Mag\kern -0.05em Up}\xspace}
\def\MagDown {\mbox{\em MagDown}\xspace}

\ifthenelse{\boolean{uprightparticles}}%
{

 \def\Ppi         {\ensuremath{\uppi}\xspace}

 \def\PDelta      {\ensuremath{\Delta}\xspace}                 
 \def\PXi         {\ensuremath{\Xi}\xspace}                 
 \def\PLambda     {\ensuremath{\Lambda}\xspace}                 
 \def\PSigma      {\ensuremath{\Sigma}\xspace}                 
 \def\POmega      {\ensuremath{\Omega}\xspace}                 
 \def\PUpsilon    {\ensuremath{\Upsilon}\xspace}
 \let\oldPi\Pi
 \def\PPi         {\ensuremath{\oldPi}\xspace}

 \def\PB      {\ensuremath{\mathrm{B}}\xspace}                 
                  
 \def\PD      {\ensuremath{\mathrm{D}}\xspace}

 \def\PK      {\ensuremath{\mathrm{K}}\xspace}

 \def\PV      {\ensuremath{\mathrm{V}}\xspace}

 \def\Pb      {\ensuremath{\mathrm{b}}\xspace}                 
 \def\Pc      {\ensuremath{\mathrm{c}}\xspace}                 
 \def\Pd      {\ensuremath{\mathrm{d}}\xspace}                 
 \def\Pe      {\ensuremath{\mathrm{e}}\xspace}

 \def\Pi      {\ensuremath{\mathrm{i}}\xspace}

 \def\Ps      {\ensuremath{\mathrm{s}}\xspace}                 
                  
 \def\Pu      {\ensuremath{\mathrm{u}}\xspace}

 \def\thebaroffset{0.0em}
}
{

 \def\Ppi         {\ensuremath{\pi}\xspace}

 \mathchardef\PDelta="7101
 \mathchardef\PXi="7104
 \mathchardef\PLambda="7103
 \mathchardef\PSigma="7106
 \mathchardef\POmega="710A
 \mathchardef\PUpsilon="7107
 \mathchardef\PPi="7105
                  
 \def\PB      {\ensuremath{B}\xspace}                 
                  
 \def\PD      {\ensuremath{D}\xspace}

 \def\PK      {\ensuremath{K}\xspace}

 \def\PV      {\ensuremath{V}\xspace}

 \def\Pb      {\ensuremath{b}\xspace}                 
 \def\Pc      {\ensuremath{c}\xspace}                 
 \def\Pd      {\ensuremath{d}\xspace}                 
 \def\Pe      {\ensuremath{e}\xspace}

 \def\Pi      {\ensuremath{i}\xspace}

 \def\Ps      {\ensuremath{s}\xspace}                 
                  
 \def\Pu      {\ensuremath{u}\xspace}

 \def\thebaroffset{0.18em}
}
\newcommand{\offsetoverline}[2][\thebaroffset]{\kern #1\overline{\kern -#1 #2}}%

\makeatletter
\ifcase \@ptsize \relax% 10pt
  \newcommand{\miniscule}{\@setfontsize\miniscule{4}{5}}% \tiny: 5/6
\or% 11pt
  \newcommand{\miniscule}{\@setfontsize\miniscule{5}{6}}% \tiny: 6/7
\or% 12pt
  \newcommand{\miniscule}{\@setfontsize\miniscule{5}{6}}% \tiny: 6/7
\fi
\makeatother

\DeclareRobustCommand{\optbar}[1]{\shortstack{{\miniscule (\rule[.5ex]{1.25em}{.18mm})}
  \\ [-.7ex] $#1$}}

\def\epem       {{\ensuremath{\Pe^+\Pe^-}}\xspace}
\def\uquark    {{\ensuremath{\Pu}}\xspace}

\def\dquark    {{\ensuremath{\Pd}}\xspace}

\def\squark    {{\ensuremath{\Ps}}\xspace}

\def\cquark    {{\ensuremath{\Pc}}\xspace}

\def\bquark    {{\ensuremath{\Pb}}\xspace}

\def\pion   {{\ensuremath{\Ppi}}\xspace}

\def\pip    {{\ensuremath{\pion^+}}\xspace}
\def\pim    {{\ensuremath{\pion^-}}\xspace}

\def\pimp   {{\ensuremath{\pion^\mp}}\xspace}

\def\kaon    {{\ensuremath{\PK}}\xspace}
\def\KorKbar {\kern \thebaroffset\optbar{\kern -\thebaroffset \PK}{}\xspace}

\def\Kp      {{\ensuremath{\kaon^+}}\xspace}
\def\Km      {{\ensuremath{\kaon^-}}\xspace}
\def\Kpm     {{\ensuremath{\kaon^\pm}}\xspace}

\def\KS      {{\ensuremath{\kaon^0_{\mathrm{S}}}}\xspace}

\def\Dbar    {{\ensuremath{\offsetoverline{\PD}}}\xspace}
\def\D       {{\ensuremath{\PD}}\xspace}

\def\DorDbar {\kern \thebaroffset\optbar{\kern -\thebaroffset \PD}\xspace}
\def\Dz      {{\ensuremath{\D^0}}\xspace}
\def\Dzb     {{\ensuremath{\Dbar{}^0}}\xspace}
\def\Dp      {{\ensuremath{\D^+}}\xspace}
\def\Dm      {{\ensuremath{\D^-}}\xspace}

\def\DpDm    {\ensuremath{\Dp {\kern -0.16em \Dm}}\xspace}

\def\Dstarp  {{\ensuremath{\D^{*+}}}\xspace}
\def\Dstarm  {{\ensuremath{\D^{*-}}}\xspace}

\def\theDstarp{{\ensuremath{\D^{*}(2010)^{+}}}\xspace}

\def\Ds      {{\ensuremath{\D^+_\squark}}\xspace}

\def\B       {{\ensuremath{\PB}}\xspace}
\def\Bbar    {{\ensuremath{\offsetoverline{\PB}}}\xspace}

\def\BorBbar {\kern \thebaroffset\optbar{\kern -\thebaroffset \PB}\xspace}
\def\Bz      {{\ensuremath{\B^0}}\xspace}
\def\Bzb     {{\ensuremath{\Bbar{}^0}}\xspace}
\def\Bd      {{\ensuremath{\B^0}}\xspace}

\def\BdorBdbar {\kern \thebaroffset\optbar{\kern -\thebaroffset \Bd}\xspace}

\def\Bs      {{\ensuremath{\B^0_\squark}}\xspace}

\def\BsorBsbar {\kern \thebaroffset\optbar{\kern -\thebaroffset \Bs}\xspace}

\def\Y#1S{\ensuremath{\PUpsilon{(#1S)}}\xspace}

\def\LorLbar     {\kern \thebaroffset\optbar{\kern -\thebaroffset \PLambda}\xspace}

\newcommand{\decay}[2]{\ensuremath{#1\!\to #2}\xspace} 

\def\to                 {\ensuremath{\rightarrow}\xspace}

\newcommand{\tauDz}{{\ensuremath{\tau_{\Dz}}}\xspace}

\def\CP                {{\ensuremath{C\!P}}\xspace}
\def\CPT               {{\ensuremath{C\!PT}}\xspace}

\def\Vud  {{\ensuremath{V_{\uquark\dquark}^{\phantom{\ast}}}}\xspace}

\def\Vus  {{\ensuremath{V_{\uquark\squark}^{\phantom{\ast}}}}\xspace}

\def\Vcds  {{\ensuremath{V_{\cquark\dquark}^\ast}}\xspace}

\def\Vcss  {{\ensuremath{V_{\cquark\squark}^\ast}}\xspace}

\def\AT#1     {\ensuremath{A_{\mathrm{T}}^{#1}}\xspace}           % 2

\def\C#1      {\ensuremath{\mathcal{C}_{#1}}\xspace}                       % 9
\def\Cp#1     {\ensuremath{\mathcal{C}_{#1}^{'}}\xspace}                    % 7
\def\Ceff#1   {\ensuremath{\mathcal{C}_{#1}^{\mathrm{(eff)}}}\xspace}        % 9  
\def\Cpeff#1  {\ensuremath{\mathcal{C}_{#1}^{'\mathrm{(eff)}}}\xspace}       % 7
\def\Ope#1    {\ensuremath{\mathcal{O}_{#1}}\xspace}                       % 2
\def\Opep#1   {\ensuremath{\mathcal{O}_{#1}^{'}}\xspace}                    % 7

\newcommand{\ket}[1]{\ensuremath{|#1\rangle}}              % {b}
\newcommand{\nospaceunit}[1]{\ensuremath{\text{#1}}}       
\newcommand{\aunit}[1]{\ensuremath{\text{\,#1}}}       
\newcommand{\tev}{\aunit{Te\kern -0.1em V}\xspace}
\newcommand{\gev}{\aunit{Ge\kern -0.1em V}\xspace}
\newcommand{\mev}{\aunit{Me\kern -0.1em V}\xspace}
\newcommand{\kev}{\aunit{ke\kern -0.1em V}\xspace}
\newcommand{\ev}{\aunit{e\kern -0.1em V}\xspace}
 
\newcommand{\mevc}{\ensuremath{\aunit{Me\kern -0.1em V\!/}c}\xspace}
\newcommand{\gevc}{\ensuremath{\aunit{Ge\kern -0.1em V\!/}c}\xspace}
\newcommand{\mevcc}{\ensuremath{\aunit{Me\kern -0.1em V\!/}c^2}\xspace}
\newcommand{\gevcc}{\ensuremath{\aunit{Ge\kern -0.1em V\!/}c^2}\xspace}
\def\cm   {\aunit{cm}\xspace}

\def\mm   {\aunit{mm}\xspace}

\def\mum  {\ensuremath{\,\upmu\nospaceunit{m}}\xspace}

\def\fb   {\ensuremath{\aunit{fb}}\xspace}
\def\invfb   {\ensuremath{\fb^{-1}}\xspace}

\def\fs   {\aunit{fs}}

\newcommand{\chisq}{\ensuremath{\chi^2}\xspace}
\newcommand{\chisqndf}{\ensuremath{\chi^2/\mathrm{ndf}}\xspace}

\def\gsim{{~\raise.15em\hbox{$>$}\kern-.85em
          \lower.35em\hbox{$\sim$}~}\xspace}
\def\lsim{{~\raise.15em\hbox{$<$}\kern-.85em
          \lower.35em\hbox{$\sim$}~}\xspace}

\def\pt         {\ensuremath{p_{\mathrm{T}}}\xspace}

\def\ptot       {\ensuremath{p}\xspace}

\def\degrees{\ensuremath{^{\circ}}\xspace}

\def\mrad{\aunit{mrad}\xspace}
\def\rad{\aunit{rad}\xspace}

\def\evtgen     {\mbox{\textsc{EvtGen}}\xspace}

\def\geant      {\mbox{\textsc{Geant4}}\xspace}

\def\photos     {\mbox{\textsc{Photos}}\xspace}

\def\pythia     {\mbox{\textsc{Pythia}}\xspace}

\def\tell1  {TELL1\xspace}
\def\ukl1   {UKL1\xspace}

\newcommand{\ie}{\mbox{\itshape i.e.}\xspace}

\newcommand{\cf}{\mbox{\itshape cf.}\xspace}

\newcommand{\vs}{\mbox{\itshape vs.}\xspace}
\newcommand{\lhcborcid}[1]{\href{https://orcid.org/#1}{\hspace*{0.1em}\raisebox{-0.45ex}{\includegraphics[width=1em]{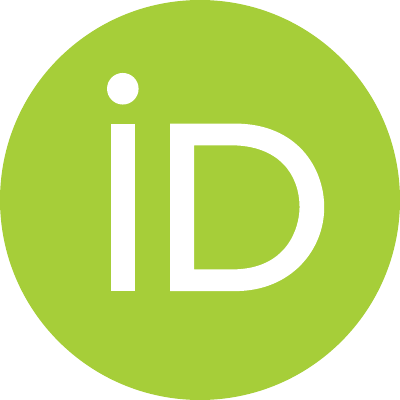}}}}

\usepackage{cite} % Allows for ranges in citations
\usepackage{mciteplus}
\usepackage{booktabs}
\usepackage{multirow}
\usepackage[capitalise]{cleveref}
\Crefname{section}{Sec.}{Secs.}
\usepackage{bm}
\newcolumntype{C}{>{$}c<{$}}
\newcolumntype{L}{>{$}l<{$}}
\newcolumntype{R}{>{$}r<{$}}

\def\spip    {{\ensuremath{\pion^{+}_{\mathrm{ \scriptscriptstyle s}}}}\xspace}

\def\suf     {{\ensuremath{\textnormal{SU(3)}_\textnormal{F}}}\xspace}
\def\af      {{\ensuremath{A_f}}\xspace}
\def\afb     {{\ensuremath{A_{\bar{f}}}}\xspace}
\def\abf     {{\ensuremath{\bar{A}_f}}\xspace}
\def\abfb    {{\ensuremath{\bar{A}_{\bar{f}}}}\xspace}

\def\DY {\ensuremath{\Delta Y}\xspace}

\def\dKpi {\ensuremath{\delta_{\kaon\pion}}\xspace}
\def\DKpi {\ensuremath{\Delta_f}\xspace}
\def\phiMKpi {\ensuremath{\phi^{M}_{f}}\xspace}
\def\phiGKpi {\ensuremath{\phi^{\Gamma}_{f}}\xspace}
\def\phiKpi {\ensuremath{\phi_{\lambda_f}}\xspace}
\def\adws {\ensuremath{a^{d}_{\text{DCS}}}\xspace}

\def\adkk {\ensuremath{a^{d}_{\kaon\kaon}}\xspace}
\def\adpipi {\ensuremath{a^{d}_{\pion\pion}}\xspace}

\def\tauDz {\ensuremath{\tau_{\Dz}}\xspace}
\def\mDz {\ensuremath{m_{\Dz}}\xspace}

\def\xprimepmsq {\ensuremath{(x^{\prime \pm})^2}\xspace}

\def\yprimepm   {\ensuremath{y^{\prime \pm}}\xspace}

\def\yprimepmsq {\ensuremath{(y^{\prime \pm})^2}\xspace}

\def\RKpip {\ensuremath{R_{\kaon\pion}^{+}}\xspace}
\def\RKpim {\ensuremath{R_{\kaon\pion}^{-}}\xspace}
\def\RKpipm {\ensuremath{R_{\kaon\pion}^{\pm}}\xspace}

\def\RKpi {\ensuremath{R_{_{\kaon\pion}}}\xspace}
\def\AKpi {\ensuremath{A_{_{\kaon\pion}}}\xspace}
\def\cKpi {\ensuremath{c_{_{\kaon\pion}}}\xspace}
\def\cKpipr {\ensuremath{c_{_{\kaon\pion}}^{\prime}}\xspace}
\def\DcKpi {\ensuremath{\Delta c_{_{\kaon\pion}}}\xspace}
\def\DcKpipr {\ensuremath{\Delta c_{_{\kaon\pion}}^{\prime}}\xspace}

\def\tAKpi {\ensuremath{\tilde{A}_{_{\kaon\pion}}}\xspace}
\def\tDcKpi {\ensuremath{\Delta \tilde{c}_{_{\kaon\pion}}}\xspace}
\def\tDcKpipr {\ensuremath{\Delta \tilde{c}_{_{\kaon\pion}}^{\prime}}\xspace}

\def\cWSp   {\ensuremath{ c_{\textnormal{WS},f}^{+}}\xspace}
\def\cWSm   {\ensuremath{ c_{\textnormal{WS},f}^{-}}\xspace}
\def\cWSpm  {\ensuremath{ c_{\textnormal{WS},f}^{\pm}}\xspace}
\def\cWSprp {\ensuremath{ c_{\textnormal{WS},f}^{\prime+}}\xspace}
\def\cWSprm {\ensuremath{ c_{\textnormal{WS},f}^{\prime-}}\xspace}
\def\cWSprpm{\ensuremath{ c_{\textnormal{WS},f}^{\prime\pm}}\xspace}

\def\PV {{\ensuremath{\text{PV}}}\xspace}

\def\IP {\ensuremath{\text{IP}}\xspace}

\def\mDp {\ensuremath{m(\Dz\spip)}\xspace}

\def\DzorDzbar {\kern \thebaroffset\optbar{\kern -\thebaroffset \D}\ensuremath{^0}\xspace}

\def\takki {\ensuremath{a_{i}^{_{\kaon\kaon}}}\xspace}
\def\AIi {\ensuremath{{A}_{i}^{\text{I}}}\xspace}

\def\Ripm {\ensuremath{R^{\pm}_{i}}\xspace}

\def\tRipm {\ensuremath{\widetilde{R}_{i}^{\pm}}\xspace}

\def\DY {\ensuremath{\Delta Y}\xspace}

\begin{document}

%%%%%%%%%%%%%%%%%%%%%%%%%
%%%%% Title     %%%%%%%%%
%%%%%%%%%%%%%%%%%%%%%%%%%
\renewcommand{\thefootnote}{\fnsymbol{footnote}}
\setcounter{footnote}{1}

% %%%%%%% CHOOSE TITLE PAGE--------
%\onecolumn
% ===============================================================================
% Purpose: LHCb-PAPER journal paper title page template
% Author: 
% Created on: 2010-09-25
% ===============================================================================

%%%%%%%%%%%%%%%%%%%%%%%%%
%%%%%  TITLE PAGE  %%%%%%
%%%%%%%%%%%%%%%%%%%%%%%%%
\begin{titlepage}
\pagenumbering{roman}

% Header ---------------------------------------------------
\vspace*{-1.5cm}
\centerline{\large EUROPEAN ORGANIZATION FOR NUCLEAR RESEARCH (CERN)}
\vspace*{1.5cm}
\noindent
\begin{tabular*}{\linewidth}{lc@{\extracolsep{\fill}}r@{\extracolsep{0pt}}}
\ifthenelse{\boolean{pdflatex}}% Logo format choice
{\vspace*{-1.5cm}\mbox{\!\!\!\includegraphics[width=.14\textwidth]{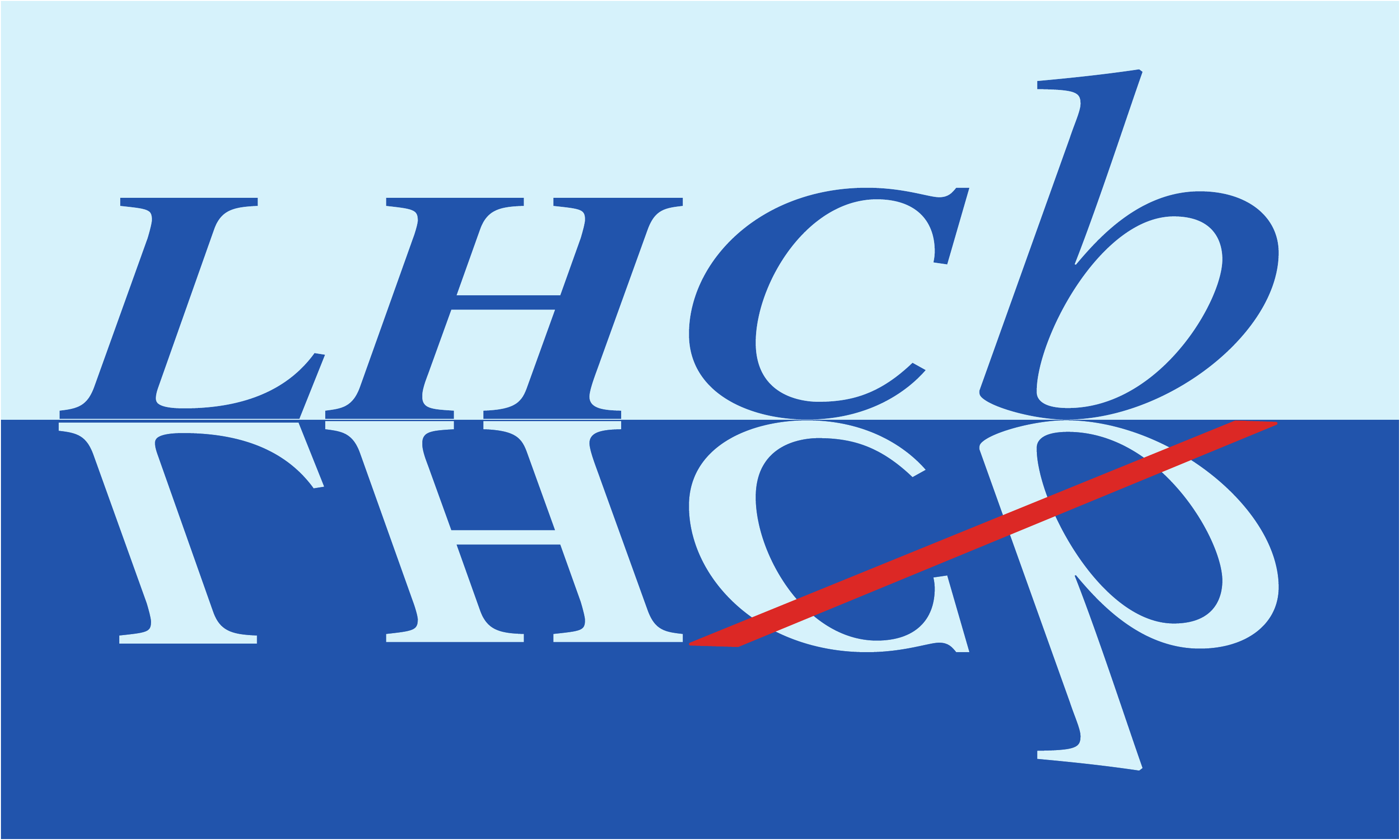}} & &}%
{\vspace*{-1.2cm}\mbox{\!\!\!\includegraphics[width=.12\textwidth]{figs/lhcb-logo.eps}} & &}%
\\
 & & CERN-EP-2024-178 \\  % ID 
 & & LHCb-PAPER-2024-008 \\  % ID 
 & & January 3, 2025 \\ % Date - Can also hardwire e.g.: 23 March 2010
 & & \\
\end{tabular*}

\vspace*{2.0cm}

% Title --------------------------------------------------
{\normalfont\bfseries\boldmath\huge
\begin{center}
% DO NOT EDIT HERE. Instead edit macro in main.tex to keep metadata correct
  \papertitle 
\end{center}
}

%\vspace*{2.0cm}
\vspace*{1cm}

% Authors -------------------------------------------------
\begin{center}
% Edit macro in main.tex to keep metadata correct
\paperauthors\footnote{Authors are listed at the end of this paper.}
\end{center}

\vspace{\fill}

% Abstract -----------------------------------------------
\begin{abstract}
  \noindent
    A measurement of the time-dependent ratio of the \decay{\Dz}{\Kp\pim} to \decay{\Dzb}{\Kp\pim} decay rates is reported.
    The analysis uses a sample of proton-proton collisions corresponding to an integrated luminosity of $6 \invfb$ recorded by the LHCb experiment from 2015 through 2018 at a center-of-mass energy of $13\tev$.
    The \Dz meson is required to originate from a \decay{\theDstarp}{\Dz\pip} decay, such that its flavor at production is inferred from the charge of the accompanying pion.
    The measurement is performed simultaneously for the \Kp\pim and \Km\pip final states, allowing both mixing and \CP-violation parameters to be determined.
    The value of the ratio of the decay rates at production is determined to be $\RKpi = (343.1 \pm 2.0) \times 10^{-5}$.  The mixing parameters are measured to be $\cKpi = (51.4 \pm 3.5) \times 10^{-4}$ and $\cKpipr = (13 \pm 4) \times 10^{-6}$, where $\sqrt{\RKpi}\cKpi$ is the linear coefficient of the expansion of the ratio as a function of decay time in units of the \Dz lifetime, and \cKpipr is the quadratic coefficient, both averaged between the \Kp\pim and \Km\pip final states.
    The precision is improved relative to the previous best measurement by approximately 60\%.
    No evidence for \CP violation is found.
\end{abstract}

\vspace*{1.0cm}

\begin{center}
  Published in Phys.~Rev.~D 111 (2025) 012001
\end{center}

\vspace{\fill}

{\footnotesize 
% Edit macro in main.tex to keep metadata correct
\centerline{\copyright~\papercopyright. \href{\paperlicenceurl}{\paperlicence}.}}
\vspace*{2mm}

\end{titlepage}

%%%%%%%%%%%%%%%%%%%%%%%%%%%%%%%%
%%%%%  EOD OF TITLE PAGE  %%%%%%
%%%%%%%%%%%%%%%%%%%%%%%%%%%%%%%%

%  empty page follows the title page ----
\newpage
\setcounter{page}{2}
\mbox{~}
%\newpage
%
%% Author List ----------------------------
%%  You need to get a new author list!
%\input{LHCb_authorlist.tex}
%
%The author list for journal publications is provided by the Membership Committee shortly after 'approval to go to paper' has been given.
%%It will be made available on the page
%%\verb!http://www.physik.uzh.ch/~strauman/forMemCo/LHCb-PAPER-XXXX-XXX/! .
%It will be sent to you by email shortly after a paper number has beens assigned.
%The author list should be included already at first circulation, 
%to allow new members of the collaboration to verify whether they have been included correctly.
%Occasionally a misspelled name is corrected or associated institutions become full members.
%In that case, a new author list will be sent to you.
%In case line numbering doesn't work well after including the authorlist, try moving the \verb!\bigskip! after the last author to a separate line.
%\twocolumn
% %%%%%%%%%%%%% ---------

\renewcommand{\thefootnote}{\arabic{footnote}}
\setcounter{footnote}{0}

%%%%%%%%%%%%%%%%%%%%%%%%%%%%%%%%
%%%%%  Table of Content   %%%%%%
%%%%%%%%%%%%%%%%%%%%%%%%%%%%%%%%
%%%% Uncomment if desired
%\tableofcontents
\cleardoublepage

%%%%%%%%%%%%%%%%%%%%%%%%%
%%%%% Main text %%%%%%%%%
%%%%%%%%%%%%%%%%%%%%%%%%%

\pagestyle{plain} % restore page numbers for the main text
\setcounter{page}{1}
\pagenumbering{arabic}

%% Uncomment during review phase. 
%% Comment before a final submission.
%\linenumbers

%% This is the main body
%% It is useful to have a single file so comemnts are not missed in overleaf.

\section{Introduction}\label{sec:introduction}
Flavor-changing neutral currents (FCNCs) and violation of charge-parity (\CP) symmetry in charm-hadron decays are strongly suppressed by the Glashow--Iliopoulous--Maiani mechanism~\cite{Glashow:1970gm,Buras:2020_GIM} and by the smallness of the elements of the Cabibbo--Kobayashi--Maskawa (CKM) matrix~\cite{Cabibbo:1963yz,Kobayashi:1973fv} connecting the first two generations of quarks with the third.
As a result, measurements of these processes can provide precise null tests of the Standard Model (SM)~\cite{EuropeanStrategyforParticlePhysicsPreparatoryGroup:2019qin}.
The ultimate precision of these tests, however, is limited by theoretical uncertainties on contributions from long-distance quantum-chromodynamics (QCD), as reviewed in  Refs.~\cite{Bediaga:2020qxg,Lenz:2020awd,Gisbert:2020vjx,Pajero:2022vev}.
Additional measurements are needed both to improve knowledge of the size of these small effects, some of which are yet to be observed, and to help clarify the magnitude of long-distance QCD contributions.

The decay \decay{\Dz(t)}{\Kp\pim}, where $\Dz(t)$ denotes the state of a neutral charm meson, at proper time $t$, that was produced in the \Dz flavor eigenstate at $t=0$, is an excellent candidate to study the effects described above (throughout this document, the inclusion of charge-conjugate processes is implied unless stated otherwise).
This process receives contributions from interfering amplitudes of comparable magnitudes from the doubly Cabibbo-suppressed \decay{\Dz}{\Kp\pim} decay and from the Cabibbo-favored \decay{\Dzb}{\Kp\pim} decay following a \Dz--\Dzb oscillation.
Therefore, analysis of the time evolution of the decay rate provides sensitivity to the mixing parameters that determine the rate of \Dz--\Dzb oscillations.
These are \mbox{$x_{12} \equiv 2 \lvert M_{12} \rvert / \Gamma$} and \mbox{$y_{12} \equiv \lvert \Gamma_{12} \rvert / \Gamma$}~\cite{Kagan:2009gb,Grossman:2009mn}, defined from the $2\times2$ effective Hamiltonian governing the time evolution of the \Dz--\Dzb system, \mbox{$\bm{H}\equiv \bm{M} - \tfrac{i}{2}\bm{\Gamma}$}, where $\Gamma$ is the \Dz decay width.
% The latest world average of these parameters is $x_{12} = (4.0 \pm 0.5) \times 10^{-3}$ and $y_{12} = (6.36 \pm 0.20) \times 10^{-3}$~\cite{LHCb-PAPER-2021-033,LHCb-CONF-2022-003}.

The dependence of the measured quantities on the value of the \Dz lifetime can be greatly reduced by taking the ratio of the \decay{\Dz(t)}{\Kp\pim} rate with that of \decay{\Dzb(t)}{\Kp\pim}, which is dominated by a Cabibbo-favored amplitude.
This and the charge-conjugate ratio are defined as 
\begin{equation}
\RKpip(t) \equiv \frac{\Gamma(\Dz(t)\to\Kp\pim)}{\Gamma(\Dzb(t)\to\Kp\pim)}\quad\text{and}\quad \RKpim(t) \equiv \frac{\Gamma(\Dzb(t)\to\Km\pip)}{\Gamma(\Dz(t)\to\Km\pip)}\,.
\end{equation}
Hereafter, the \Dz decay time is expressed in units of \tauDz, the \Dz lifetime~\cite{PDG2022}, to keep the notation more compact.
Expanding these ratios up to second order in the small mixing parameters $x_{12}$ and $y_{12}$ and, consequently, also in $t$, one obtains
\begin{equation}
\label{eq:ratio-parametrisation}
    \RKpipm(t) \approx \RKpi(1 \pm \AKpi) + \sqrt{\RKpi(1 \pm \AKpi)} (\cKpi\pm\DcKpi)\,t + (\cKpipr\pm\DcKpipr) \,t^2\,,
\end{equation}
%where the following relations hold~\cite{Kagan:2020vri,Pajero:2021jev},
where~\cite{Kagan:2020vri,Pajero:2021jev}
\begingroup\allowdisplaybreaks
\begin{align}
    \label{eq:R}
    \RKpi &\equiv \dfrac{1}{2}\bigg(\left\lvert\frac{\afb}{\abfb}\right\rvert^2 + \left\lvert\frac{\abf}{\af}\right\rvert^2 \bigg)\,, \\
    \label{eq:A}
    \AKpi &\equiv \frac{\left\lvert\afb / \abfb\right\rvert^2 - \left\lvert\abf/\af\right\rvert^2}{\left\lvert\afb / \abfb\right\rvert^2 + \left\lvert\abf/\af\right\rvert^2}\approx \adws\,, \\
    \label{eq:c}
    \cKpi  &\approx y_{12} \cos{\phiGKpi}\cos{\DKpi} + x_{12}\cos{\phiMKpi}\sin{\DKpi}\,, \\
    \label{eq:Dc}
    \DcKpi &\approx x_{12} \sin{\phiMKpi}\cos{\DKpi} - y_{12}\sin{\phiGKpi}\sin{\DKpi}\,, \\
    \label{eq:cpr}
    \cKpipr &\approx \dfrac{1}{4} \left(x_{12}^2+y_{12}^2\right)\,,\\
    \label{eq:Dcpr}
    \DcKpipr &\approx \dfrac{1}{2}x_{12}y_{12}\sin(\phiMKpi-\phiGKpi)\,.
\end{align}
\endgroup
Here, \af (\abf) denotes the decay amplitude of a \Dz (\Dzb) meson into the final state $f = \Km\pip$; \afb (\abfb) denotes the analogous amplitude for the final state $\bar{f} = \Kp\pim$; \adws is the \CP asymmetry in doubly Cabibbo-suppressed \decay{\Dz}{\Kp\pim} decays; and the weak phases \phiMKpi, \phiGKpi, and the strong phase \DKpi are defined following the convention of Ref.~\cite{Kagan:2020vri},
\begin{equation}
    \label{eq:phase-definition}
    \begin{aligned}
        \phi_f^M      - \Delta_f &\equiv \arg(-M_{12}      \, \af / \abf)\,,\qquad
        &\phi_f^M      + \Delta_f &\equiv \arg(-M_{12}      \, \afb / \abfb)\,, \\
        \phi_f^\Gamma - \Delta_f &\equiv \arg(-\Gamma_{12} \, \af / \abf)\,,\qquad
        &\phi_f^\Gamma + \Delta_f &\equiv \arg(-\Gamma_{12} \, \afb / \abfb)\,.
    \end{aligned}
\end{equation}
Finally, \CP violation in Cabibbo-favored decays and relative corrections of the order of \RKpi are neglected in \cref{eq:c,eq:Dc,eq:cpr,eq:Dcpr}.
Compared to other conventions~\cite{BaBar:2007kib,Belle:2014yoi,LHCb-PAPER-2016-033,LHCb-PAPER-2017-046}, the parametrization of \cref{eq:ratio-parametrisation} has the advantage of distinguishing between \CP-even observables (\RKpi, \cKpi and \cKpipr) and \CP-odd observables (\AKpi, \DcKpi and \DcKpipr), where the latter vanish in the case of no \CP violation, when both \phiMKpi and \phiGKpi are equal to zero. Further details on different conventions and parametrizations are reported in \cref{app:alternative_param}.

The \CP asymmetry \AKpi provides a rigorous null test of the SM.
Since the \decay{\cquark}{\uquark\dquark\bar{s}} transition does not receive contributions from electroweak-loop (penguin) or chromomagnetic-dipole operators, any signs of \CP asymmetry in the decay larger than $10^{-5}$ would provide unambiguous evidence of new interactions~\cite{Bergmann:1999pm,Grossman:2006jg}.
The same argument applies to the Cabibbo-favored \decay{\Dz}{\Km\pip} decays, where the assumption of negligible \CP asymmetry is made since the contribution from the SM amplitude is much larger than that of doubly Cabibbo-suppressed \decay{\Dz}{\Kp\pim} decays.
%The same argument applies to the Cabibbo-favored \decay{\Dz}{\Km\pip} decays, where however the contribution from the SM amplitude is much larger than that of doubly Cabibbo-suppressed \decay{\Dz}{\Kp\pim} decays, motivating the assumption of negligible \CP asymmetry in Cabibbo-favored \decay{\Dz}{\Km\pip} decays.
The parameters \cKpi and, with lesser sensitivity, \cKpipr constrain the values of the mixing parameters $x_{12}$ and $y_{12}$ as well as the phase \DKpi.
This phase is zero in the limit of \suf flavor symmetry.
Direct determinations of \DKpi at $\epem \to \psi(3770)$ charm factories are available~\cite{CLEO:2012fel,BESIII:2022qkh} but more precise measurements would be desirable. An indirect determination, which includes these measurements as well as \lhcb measurements of the angle $\gamma$ of the CKM unitarity triangle~\cite{LHCb-PAPER-2020-019,LHCb-PAPER-2020-036,LHCb-PAPER-2022-017} and of charm mixing~\cite{LHCb-PAPER-2016-033,LHCb-PAPER-2018-038,LHCb-PAPER-2019-001,LHCb-PAPER-2021-009,LHCb-PAPER-2021-041}, achieved the best precision, yielding $\DKpi = (-10.2 \pm 2.8)\degrees$~\cite{LHCb-PAPER-2021-033,LHCb-CONF-2022-003}.
Given the large uncertainty on the small value of \DKpi, and the level of precision with which the mixing parameters are known, \mbox{$x_{12} = (4.0 \pm 0.5) \times 10^{-3}$} and \mbox{$y_{12} = (6.36 \pm 0.20) \times 10^{-3}$}~\cite{LHCb-CONF-2022-003}, an improved determination of \cKpi would mostly improve the precision on the phase \DKpi.
This would improve the knowledge of \suf breaking and of rescattering at the energy scale of the charm mass~\cite{Chau:1993ec,Buccella:1994nf,Browder:1995ay,Falk:1999ts,Gao:2006nb,Buccella:2019kpn}, which currently limits predictions of the size of \CP violation in \decay{\Dz}{\Kp\Km} and \decay{\Dz}{\pip\pim} decays~\cite{Franco:2012ck,Khodjamirian:2017zdu,Chala:2019fdb,Grossman:2019xcj,Schacht:2021jaz,Pich:2023kim,Gavrilova:2023fzy,Lenz:2023rlq}.
Measurements of \RKpi, which in the \suf limit equals $\lvert \Vcds\Vus / \Vcss\Vud \rvert^2\approx (\tan\theta_{\rm C})^4$, where $\theta_{\rm C}$ is the Cabibbo angle, can also provide better insights on the size of \suf breaking.
Moreover, the \decay{\Dz}{\Kp\pim} and \decay{\Dz}{\Km\pip} decays are a simpler system than \decay{\Dz}{\Kp\Km} and \decay{\Dz}{\pip\pim} for these studies because the distinct final-state hadrons cannot rescatter into one another.
Finally, since the contribution from \CP violation in decay is expected to be negligible in the SM, the parameters \DcKpi and, again with more limited sensitivity, \DcKpipr provide a clean measurement of \CP violation in the \Dz mixing amplitudes.
In fact, the phases \phiMKpi and \phiGKpi differ from the intrinsic mixing phases of charm mixing, $\phi^M_2$ and $\phi^\Gamma_2$, by $\mathcal{O}(10^{-6}) \rad$ (see Secs.~IV.B and IV.C.2 of Ref.~\cite{Kagan:2020vri}).
While \CP violation in charm decays has been observed in $\Delta C = 1$ amplitudes~\cite{LHCb-PAPER-2019-006,LHCb-PAPER-2022-024}, all searches for \CP violation in the mixing amplitudes to date have yielded null results~\cite{LHCb-PAPER-2016-033,LHCb-PAPER-2019-001,LHCb-PAPER-2019-032,LHCb-PAPER-2020-045,LHCb-PAPER-2021-009,LHCb-PAPER-2022-020}.
The small value of \DKpi, along with the similar sizes of $x_{12}$ and $y_{12}$, implies better sensitivity to \phiMKpi than to \phiGKpi.

This article presents a measurement of $\RKpipm(t)$ performed with proton-proton ($pp$) collision data collected by the \lhcb experiment at a center-of-mass energy of $13\tev$ from 2015 through 2018, corresponding to an integrated luminosity of $6\invfb$.
The \Dz meson is required to originate from strong \decay{\theDstarp}{\Dz\spip} decays, such that its flavor at production is determined by the charge of the low-momentum ``soft'' pion, \spip.
Hereafter the \theDstarp meson is referred to as \Dstarp.
The decay \decay{\Dz(t)}{\Kp\pim} is referred to as wrong sign (WS), as the charge of the pion is opposite to that of the soft pion from the \Dstarp decay, while the decay \decay{\Dz(t)}{\Km\pip} is referred to as right sign (RS).
The results supersede those of a previous measurement based on the subset of data collected during 2015 and 2016~\cite{LHCb-PAPER-2017-046}. The results are finally combined with those of Run~1, also taken from Ref.~\cite{LHCb-PAPER-2017-046}.

\section{Measurement overview}
\label{sec:overview}
This section introduces the methodology used in the analysis, with detailed information provided in the following sections. The analysis strategy has been revisited and improved in all aspects compared to that of Ref.~\cite{LHCb-PAPER-2017-046}, which used the data sample collected during the \lhc Run~1 (2011--2012) and the first two years (2015--2016) of the \lhc Run~2 data-taking period. The measurement described in this article extends the analysis to the full Run~2 data sample by adding data collected during 2017 and 2018, with a reanalysis of 2015 and 2016 data. The size of the full Run~2 data sample is a factor of three larger than the partial Run~2 dataset used in the previous analysis.

Signal candidates are reconstructed relying on the \decay{\Dstarp}{\Dz (\to \Kpm\pimp) \spip} decay chain.
Since the \Dz and \spip mesons are produced with nearly collinear momenta in the laboratory frame, the resolution of the \Dstarp decay-vertex position along its momentum direction is comparable with the average \Dz flight distance, approximately $1\cm$, inducing a poor decay-time resolution. The latter improves on average to about $400\mum$ if a kinematic fit~\cite{Hulsbergen:2005pu} is performed requiring the \Dz meson to originate from the primary $pp$ collision vertex (\PV), exploiting the fact that most \Dstarp mesons are produced at the \PV and decay strongly with negligible flight distance. This corresponds to an uncertainty on the measured \Dz decay time of about $0.1\,\tauDz$. This constraint also improves the \Dstarp invariant-mass resolution by a factor of two.

The dataset is divided into disjoint subsamples referred to as bins.
The division is determined by the \Dz final state (\Kp\pim and \Km\pip), data-taking period (2015--2016, 2017, 2018) and \Dz decay time.
In particular, there are 18 decay-time intervals in the range $[0.4,8.0]\;\tauDz$, chosen to be equally populated except for the last four bins, which have half the number of candidates as the other bins.
The raw WS-to-RS yield ratios and the average \Dz decay time and the average of its square are determined in each subsample by discriminating the signal from the background thanks to the distinctive shape of each component in the \Dstarp invariant-mass distribution.
Both the RS and WS signal decays have an approximately Gaussian distribution for the \Dstarp invariant mass, with a standard deviation of approximately $0.3\mevcc$.
The main background, referred to as combinatorial, consists of correctly reconstructed \Dz decays associated with an unrelated pion from the same $pp$ collision.
Another background, referred to as ghost, arises from ghost soft pions. These are fake tracks generated by combining a track stub in the vertex detector with another track stub produced by a different particle in the tracking stations downstream of the magnet (see next section for additional information on the detector layout).

The raw determinations of WS-to-RS ratios in each subsample, and for different \Dz final states, \Kp\pim and \Km\pip, are corrected for the known sources of possible bias.
The most relevant ones are the contamination of doubly misidentified \Dz decays, and the removal of common candidates from the sample of WS decays. 
Doubly misidentified \Dz decays are proper RS decays which are misreconstructed as WS decays, where a genuine kaon is misidentified as a pion and vice-versa. 
The common candidates are, instead, WS \Dstarm candidates where the same \Dz candidate is also used to reconstruct a RS \Dstarp candidate with an invariant mass within $0.9\mevcc$ (approximately three times the mass resolution) from the known \Dstarp mass value~\cite{PDG2022}.
The WS \Dstarp candidate is discarded since the RS \Dstarm candidate has a much higher probability of being genuine due to the much larger production rate and purity. This requirement also removes a small fraction of proper WS decays, potentially biasing the determination of the WS-to-RS ratio in each time bin. 

The measured ratios are biased by instrumental charge asymmetries which shift the true ratio values of the different final states (\Kp\pim and \Km\pip) in opposite directions, mimicking a \CP-violating effect. 
The residual instrumental asymmetry, $A^{I}$, that affects the measurement mainly originates from the combined effect of the soft-pion detection asymmetry and from the different probability of producing \Dstarp and \Dstarm mesons in the $pp$ collisions.
The detection efficiency of the kaon-pion pair instead cancels out in the ratio at first order, thanks to the removal of kinematic regions of the soft-pion phase space with very high charge asymmetry.
The instrumental asymmetry is determined with high precision on data, by using an abundant calibration sample of promptly produced \decay{\Dstarp}{\Dz \pip} decays, where the \Dz decays into a pair of charged kaons.

A small fraction of candidates originates from the decays of long-lived $b$ hadrons (secondary decays), biasing the decay-time determination towards higher values. 
This bias is determined and corrected by exploiting simulated samples appropriately tuned to the data, and data samples where it is possible to isolate pure secondary \Dstarp decays.
Other minor sources of decay-time biases are also considered and appropriately handled, such as biases on the measurement of the flight distance of the \Dz meson, and biases due to the misassociation of the \Dz meson to the correct reconstructed \PV.   

The expected time-dependent behavior is fitted to the measured ratios and average decay times to measure mixing and \CP-violation parameters. The fit function includes the bias sources and contributions from other systematic uncertainties with constraints applied via nuisance parameters. 
The best-fit values of the parameters of interest were examined only after the analysis strategy was finalized, to avoid experimenter bias.
The robustness and reliability of the analysis methodology are extensively tested 
looking for possible unexpected variations of the measured parameters as a function of different observables related to the kinematics and topology of the decay or different conditions of the detector, suitably chosen to be sensitive to the main sources of systematic uncertainties. 

\section{\lhcb detector}
\label{sec:detector}

The \lhcb detector~\cite{LHCb-DP-2008-001,LHCb-DP-2014-002} is a single-arm forward spectrometer covering the pseudorapidity range $2<\eta <5$, designed for the study of particles containing \bquark or \cquark quarks. The detector includes a high-precision tracking system consisting of a silicon-strip vertex detector surrounding the $pp$ interaction region, a large-area silicon-strip detector located upstream of a dipole magnet with a bending power of about $4{\mathrm{\,T\,m}}$, and three stations of silicon-strip detectors and straw drift tubes placed downstream of the magnet.
The tracking system provides a measurement of the momentum, \ptot, of charged particles with a relative uncertainty that varies from 0.5\% at low momentum to 1.0\% at 200\gevc.
The minimum distance of a track to a PV, the impact parameter (\IP), is measured with a resolution of $(15+29/\pt)\mum$, where \pt is the component of the momentum transverse to the beam, in\,\gevc.
The magnetic field deflects oppositely charged particles in opposite directions, which can lead to detection asymmetries.
Therefore, its polarity is reversed around every two weeks throughout the data taking to reduce such effects.
Different types of charged hadrons are distinguished using information from two ring-imaging Cherenkov (\rich) detectors.
Photons, electrons and hadrons are identified by a calorimeter system consisting of scintillating-pad and preshower detectors, an electromagnetic and a hadronic calorimeter. Muons are identified by a system composed of alternating layers of iron and multiwire proportional chambers.

The online event selection is performed by a trigger, which consists of a hardware stage followed by a two-level software stage, which applies a full event reconstruction.
At the hardware-trigger stage, events are required to contain a muon with high \pt or a hadron, photon or electron with high transverse energy deposited in the calorimeters.
For hadrons, the transverse energy threshold is approximately $3.7\gev$.
In between the two software stages, an alignment and calibration of the detector are performed in near real-time~\cite{LHCb-PROC-2015-011} and updated constants are made available for the trigger, ensuring high-quality tracking and particle identification (PID) information.
The excellent performance of the online reconstruction
offers the opportunity to perform physics analyses directly using candidates reconstructed at the trigger level~\cite{LHCb-DP-2012-004,LHCb-DP-2016-001}, which the present analysis exploits.
The storage of only the triggered candidates enables a reduction in the event size by an order of magnitude.

Simulation is used to estimate the size of the bias in the measurement of the \Dz decay time, as described in \cref{sec:time_bias}.
In the simulation, $pp$ collisions are generated using \pythia~\cite{Sjostrand:2007gs,*Sjostrand:2006za} with a specific \lhcb configuration~\cite{LHCb-PROC-2010-056}.
Decays of unstable particles are described by \evtgen~\cite{Lange:2001uf}, in which final-state radiation is generated using \photos~\cite{davidson2015photos}.
The interaction of the generated particles with the detector, and its response, are implemented using the \geant toolkit~\cite{Allison:2006ve, *Agostinelli:2002hh} as described in Ref.~\cite{LHCb-PROC-2011-006}. In order to increase the speed of producing simulated events and allow much larger samples to be saved on disk, the simulation of a single specific decay process can be enabled, neglecting all other particles emerging from the $pp$ collisions. More details on the treatment of these samples are given in \cref{sec:time_bias}.

\section{Candidate selection}\label{sec:selection}
The \decay{\Dstarp}{\Dz}{\spip} decay, where the \Dz meson decays into the $\Kp\pim$ or $\Km\pip$ final state, is fully reconstructed online and selected by a dedicated trigger.
The hardware trigger requires large transverse energy deposited in the calorimeters by one or both of the \Dz decay products or, alternatively, the hardware-trigger decision can be independent of the \Dz decay products and the soft pion.
This requirement discards candidates selected because of the energy deposit of the soft pion and avoids the introduction of additional detection charge asymmetries.
At least one (or both) of the tracks from the \Dz decay must meet the selection criteria of the single-track (two-track) first-stage software trigger.
The former requires the presence of at least one track with high \pt and high IP significance with respect to all PVs.
The latter requires that two high-\pt tracks form a good-quality vertex that is significantly displaced from its associated PV, defined as the PV to which the IP significance of the two-track combination is the smallest.
This selection is based on a boosted decision tree classifier~\cite{BBDT} that takes as inputs the \chisq of the two-track vertex fit, the number of tracks with high IP significance with respect to the PV ($p\textnormal{-value}<3.3\times10^{-4}$), the sum of the \pt of the two tracks, and the significance of the flight distance of their combination from the associated PV.
The second-stage software trigger combines high-quality tracks with opposite charges and a distance of closest approach smaller than $0.1\mm$ into \Dz candidates.
Each track is required to have $p > 5 \gevc$, $\pt > 800 \mevc$ and is assigned a pion or kaon mass hypothesis based on information from the \rich detectors.
The \Dz decay vertex is required to be significantly displaced from the PV, and the angle between the \Dz momentum and the vector connecting the PV and the \Dz decay vertex is required to be less than one degree. The \pt of the \Dz meson is also required to be higher than 1\gevc.  Finally, all high-quality tracks in the event that form a good-quality vertex with the \Dz candidate, are identified as soft pions and are used to form \Dstarp candidates.

In the offline selection, \spip candidates are required not to be identified as electrons or kaons by the \rich detectors and calorimeters, and are required to satisfy $\pt(\spip) > 200\mevc$ and $\eta(\spip)<4.3$ to reduce the combinatorial background from random associations of \Dz mesons with unrelated tracks. The requirement on the pseudorapidity also reduces detection charge asymmetries due to material interactions in the conical beampipe of the RICH detector placed after the vertex detector.
The \mDp observable is the \Dstarp invariant mass measured by fixing the \Dz mass to its known value~\cite{PDG2022} when evaluating the \Dstarp candidate energy, in order to reduce the contribution from the \Dz invariant-mass resolution.

\begin{figure}[tb]
    \centering
    \includegraphics[width=0.47\linewidth]{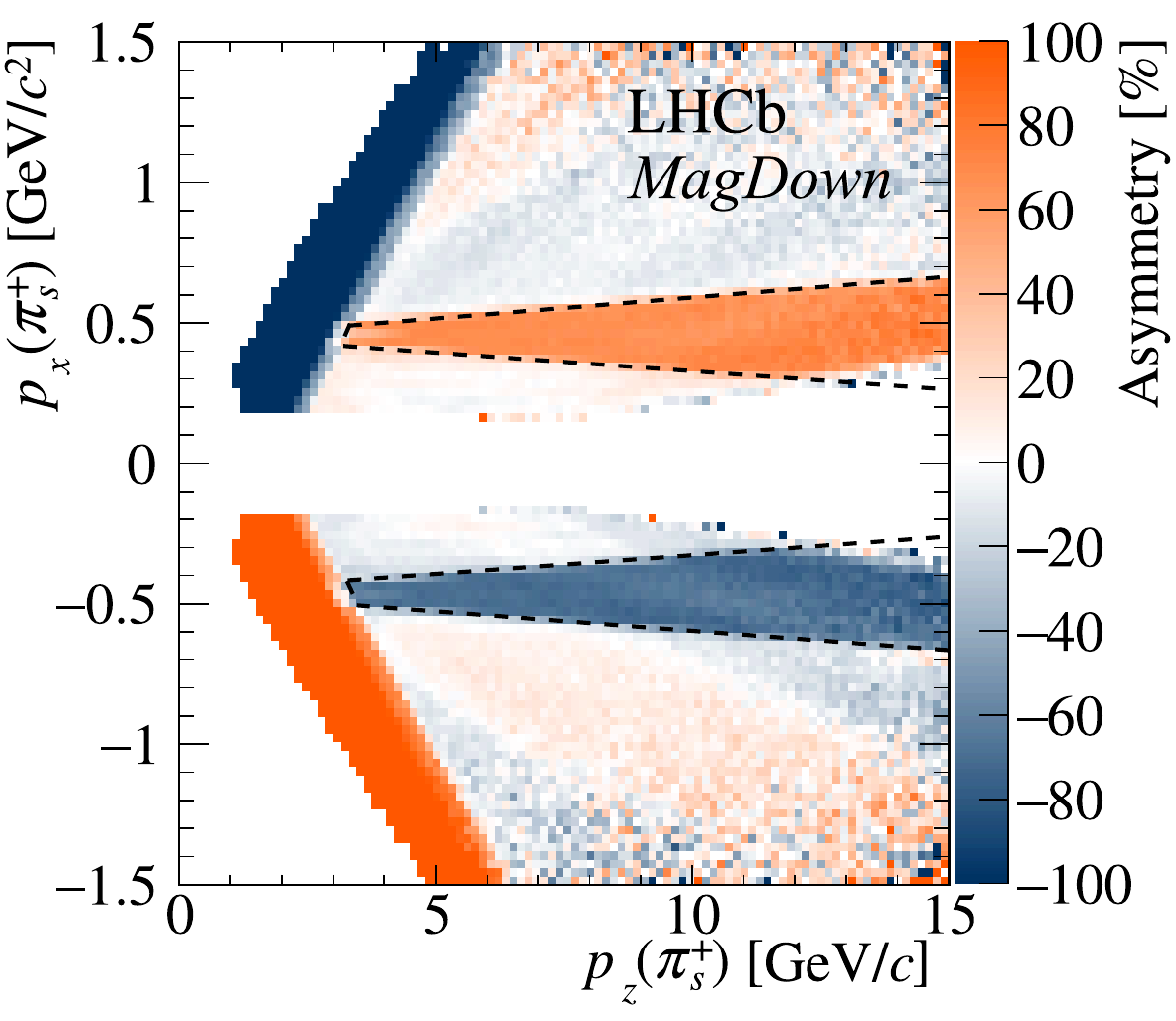}\qquad
    \includegraphics[width=0.47\linewidth]{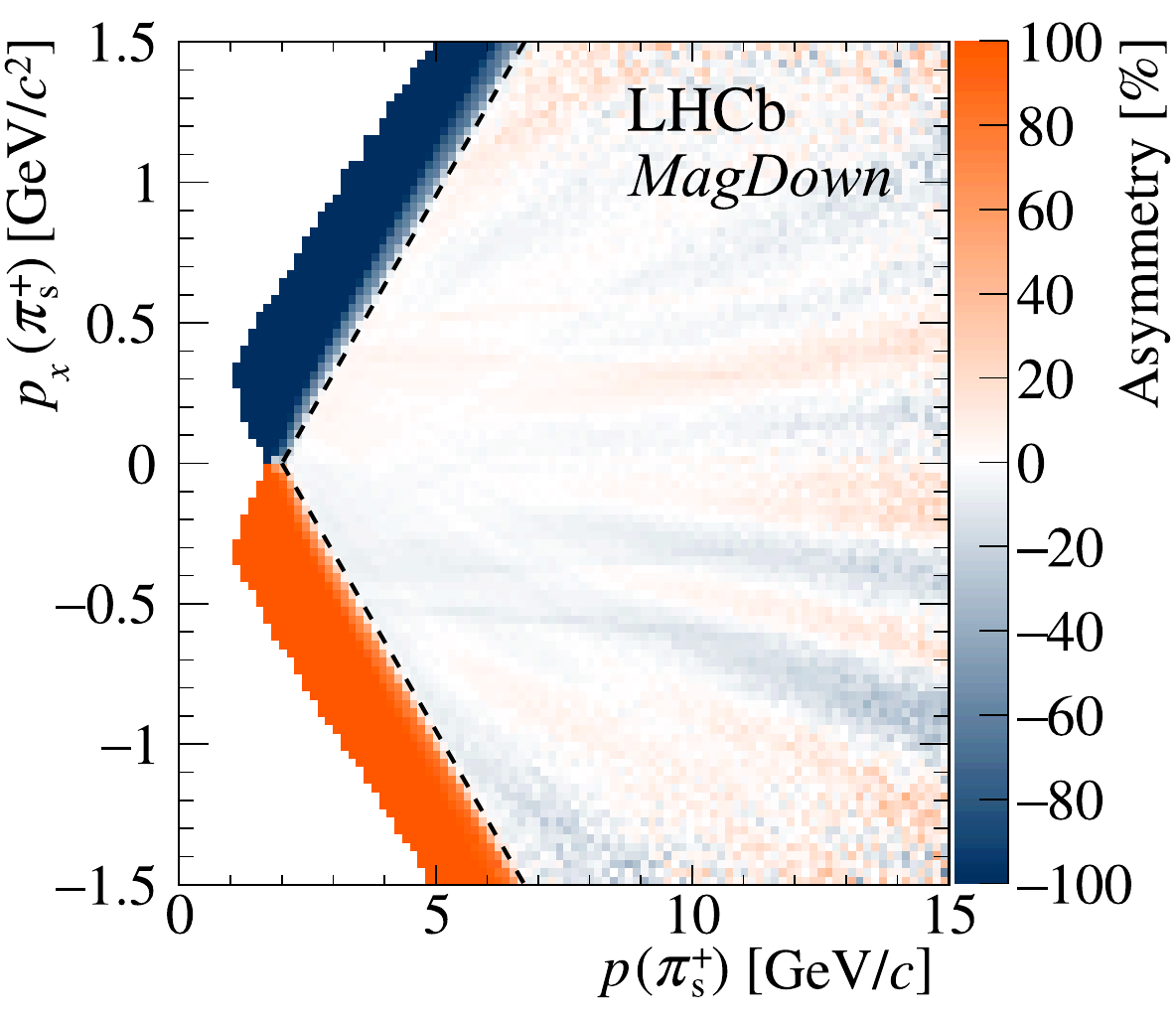}
    \caption{Raw asymmetries of the soft pion for the RS signal candidates, with the polarity of the magnet pointing towards the negative direction of the $y$ axis (\MagDown). Dashed lines show the excluded regions of high detection asymmetries. The left plot includes only candidates with $|p_y/p_z| < 0.015$, \ie the kinematic region close to the beam pipe. The right plot shows all candidates, except those excluded by the beam-pipe fiducial requirements.}
    \label{fig:fiducial}
\end{figure}

Whenever multiple \Dstarp candidates are formed using the same \Dz candidate and different soft pions with the same charge, these \Dstarp candidates are discarded, since half or more of them are background.
This happens in around 9\% of WS candidates.
In addition, whenever a WS candidate is reconstructed also as a RS candidate, and the \mDp value of the RS candidate lies within $\pm 0.9\mevcc$ of the known \Dstarp mass~\cite{PDG2022}, the WS candidate is discarded since it is likely to be either combinatorial or ghost background.
This requirement rejects around 16\% of the WS sample.
About 40\% of the residual ghost background is removed based on the output of a multivariate classifier trained using information from all tracking subdetectors to differentiate between genuine and fake tracks~\cite{DeCian:2255039}.
The soft pion is also required to satisfy the following condition, where momenta are expressed in\,\mevc:
\begin{align}\label{eq:fiducial}
|p_x| &< 0.317\cdot(p-2000)
\;\text{AND} \nonumber\\
&\left[\left|\frac{p_y}{p_z}\right| > 0.015
\;\text{OR}\; |p_x| < 470 -0.01397\cdot p_z
\;\text{OR}\; |p_x| > 430 + 0.01605 \cdot p_z\right]\,,
\end{align}
where $p_x$, $p_y$, and $p_z$ are the components of the soft-pion momentum projected onto the \lhcb coordinate system.\interfootnotelinepenalty=10000\footnote{The LHCb coordinate system is a right-handed system centered on the nominal $pp$ collision point, with the $z$ axis pointing along the beam direction towards the detectors downstream, the $y$ axis pointing vertically upwards, and the $x$ axis pointing in the horizontal direction.}
This requirement rejects about 15\% of the signal candidates and an additional 20\% of the ghost background while also removing candidates prone to large detector asymmetries, as shown in \cref{fig:fiducial}. This approach closely follows that used in other \lhcb precision measurements~\cite{LHCb-PAPER-2019-006,LHCb-PAPER-2022-024}, and ensures precise and robust cancellation of instrumental charge asymmetries in the WS-to-RS yield ratios, as described in \cref{sec:asymmetry_bias}.
The residual background from \Dz decays with a single misidentified decay product is further strongly reduced by requiring the \Dz invariant mass, $m(K\pi)$, to be within $\pm 24\mevcc$ of the \Dz mass~\cite{PDG2022}, corresponding to around three times the mass resolution.
The background from \Dz decays where the mass assignment of both the decay products is inverted is reduced by computing the \Dz invariant mass with swapped kaon-pion mass hypotheses, $m(K\pi)_{\textrm{swap}}$, and requiring it to be more than $16\mevcc$ away from the \Dz mass.
This requirement removes about 11\% of the signal sample while rejecting 80\% of the doubly misidentified background.
Secondary \Dstarp decays, unlike promptly produced decays, do not always point back to the \PV.
Their fraction is reduced to a few percent by requiring $\IP(\Dz) < 60\mum$.

\begin{figure}
    \centering
    \includegraphics[width=0.48\linewidth]{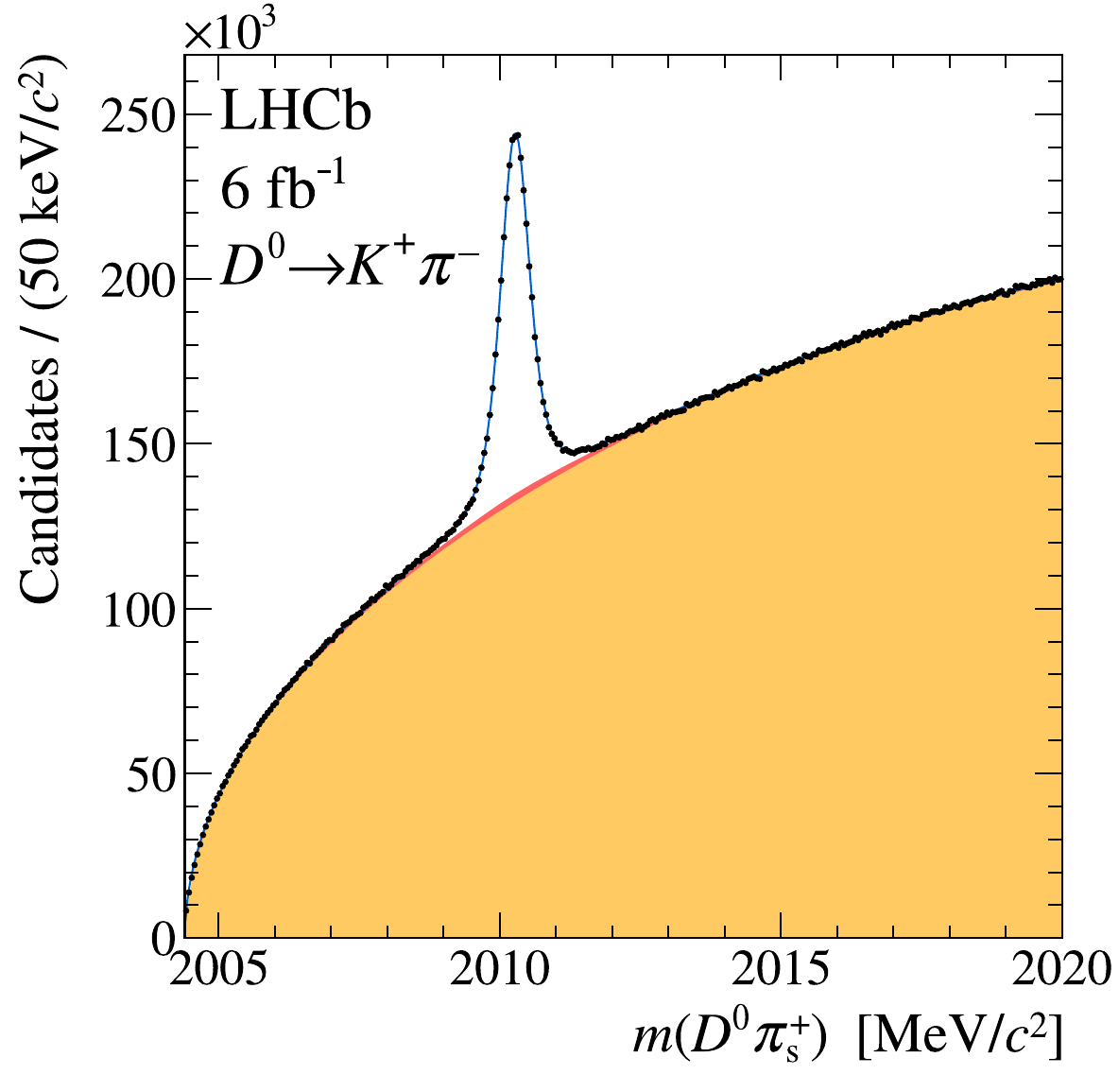}
    \includegraphics[width=0.48\linewidth]{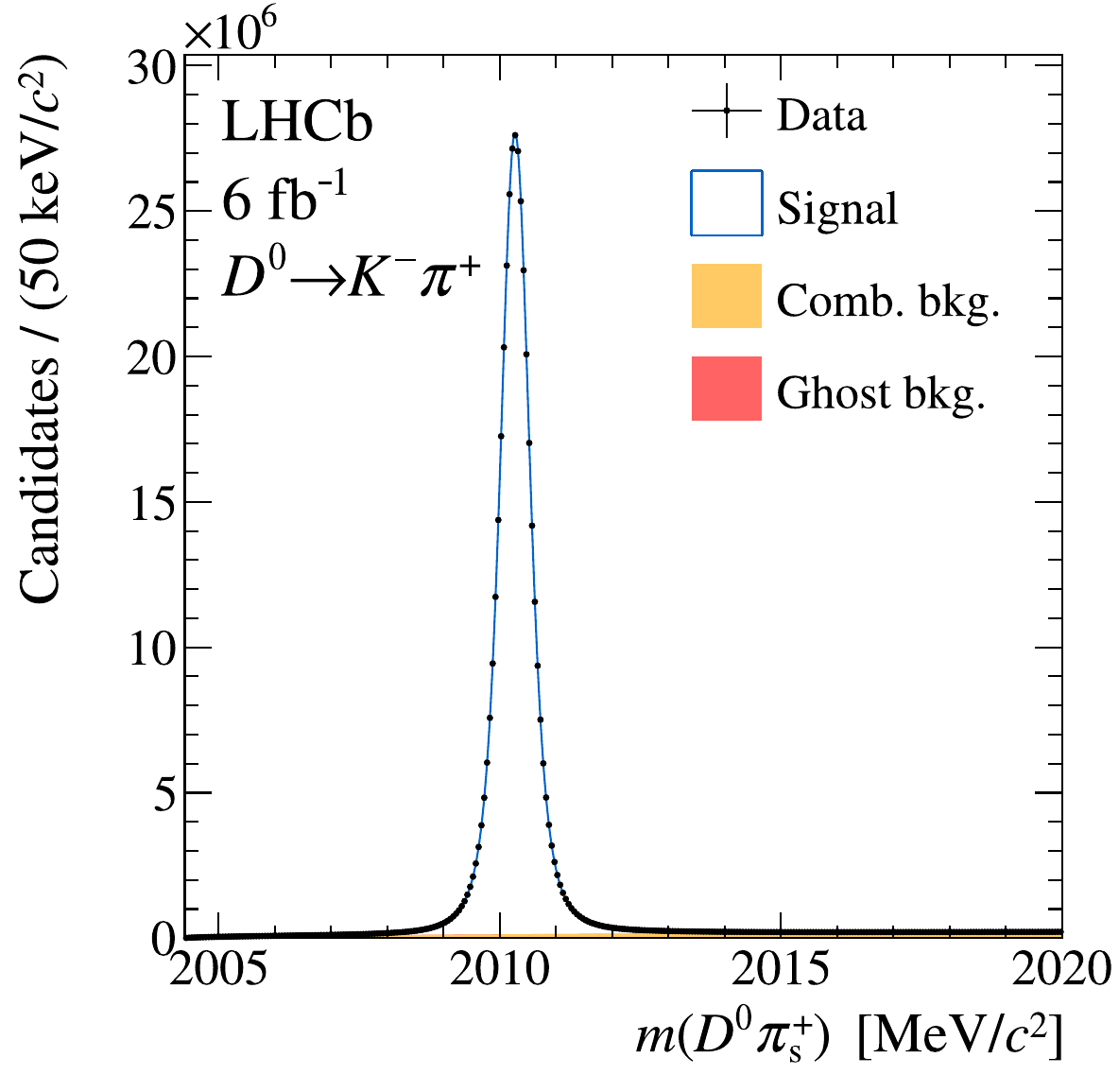}
    \caption{Mass distribution of (left) WS and (right) RS candidates after the offline selection. Different fit components are displayed stacked. The ghost background component is present in both WS and RS samples, but is barely visible only in the WS one.
    }
    \label{fig:final_sample}
\end{figure}

The \mDp distributions of WS and RS candidates after all selections are shown in Fig.~\ref{fig:final_sample}.
%%%%%%%%%%%
%%%%%%%%%%%
The result of a fit to the \mDp distribution of the whole data sample is superimposed.
The probability density functions (PDFs) employed in the fit are described in \cref{sec:mass_fit}.
Signal decays are distributed according to an approximately Gaussian PDF in the \Dstarp invariant mass with a standard deviation of about 0.3\mevcc.
The background is dominated by real \Dz mesons associated with uncorrelated particles and has a square-root-like shape.
The small contribution from the ghost soft-pion background accounts for about 3.7\% of the whole WS sample and will be discussed and quantified in the next two sections.
Selecting a mass range within $\pm 0.6\mevcc$ (about two standard deviations) from the known \Dstarp mass, purities of about 99.1\% and 30.9\% are found for the RS and WS candidates, respectively. 
The signal yield is 1.6 million for WS decays and 412 million for RS decays.

\section{Ghost background sample}
\label{sec:CG}

\begin{figure}[tb]
\begin{center}
    \includegraphics[width=0.485\linewidth]{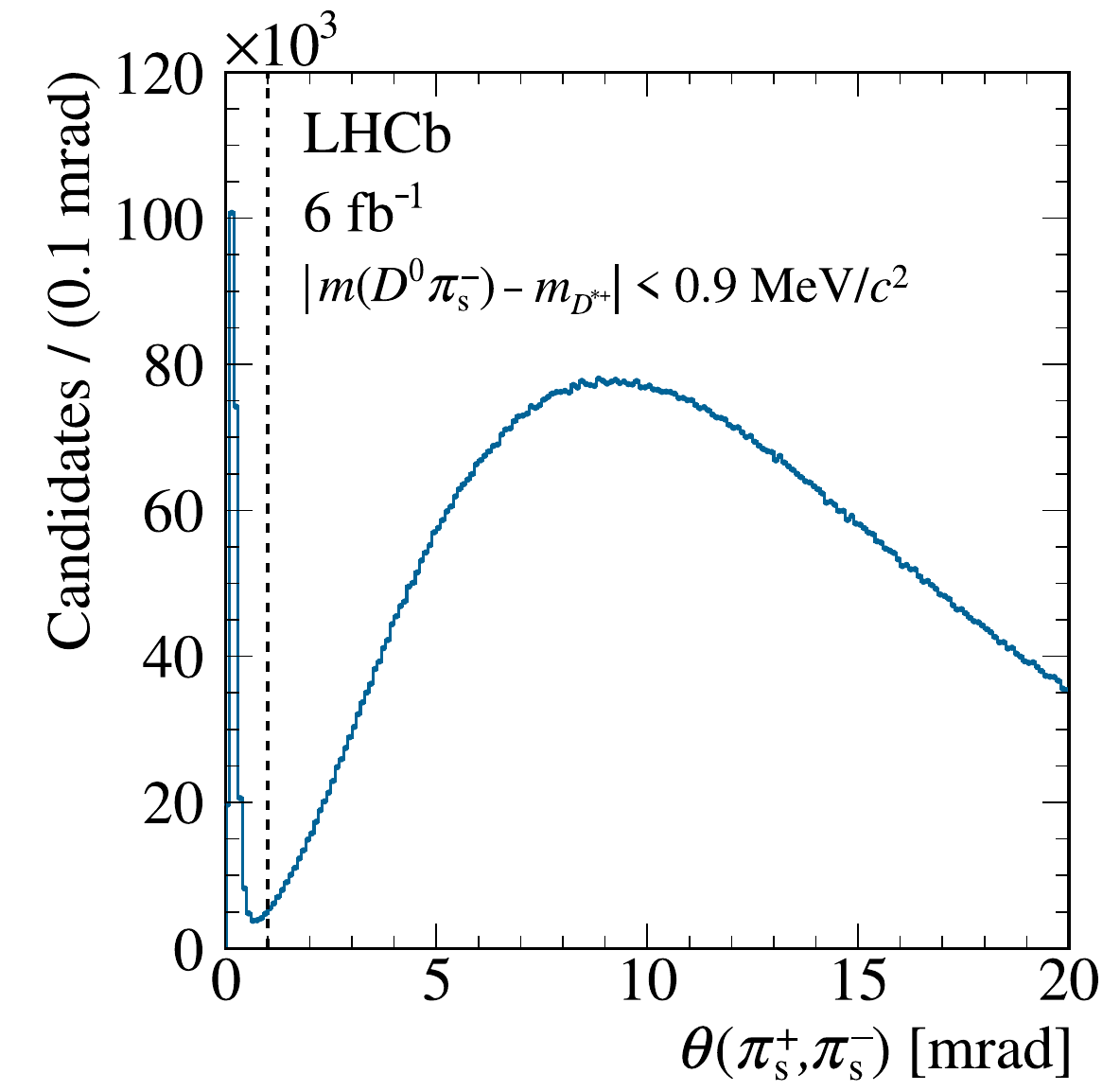}
    \includegraphics[width=0.485\linewidth]{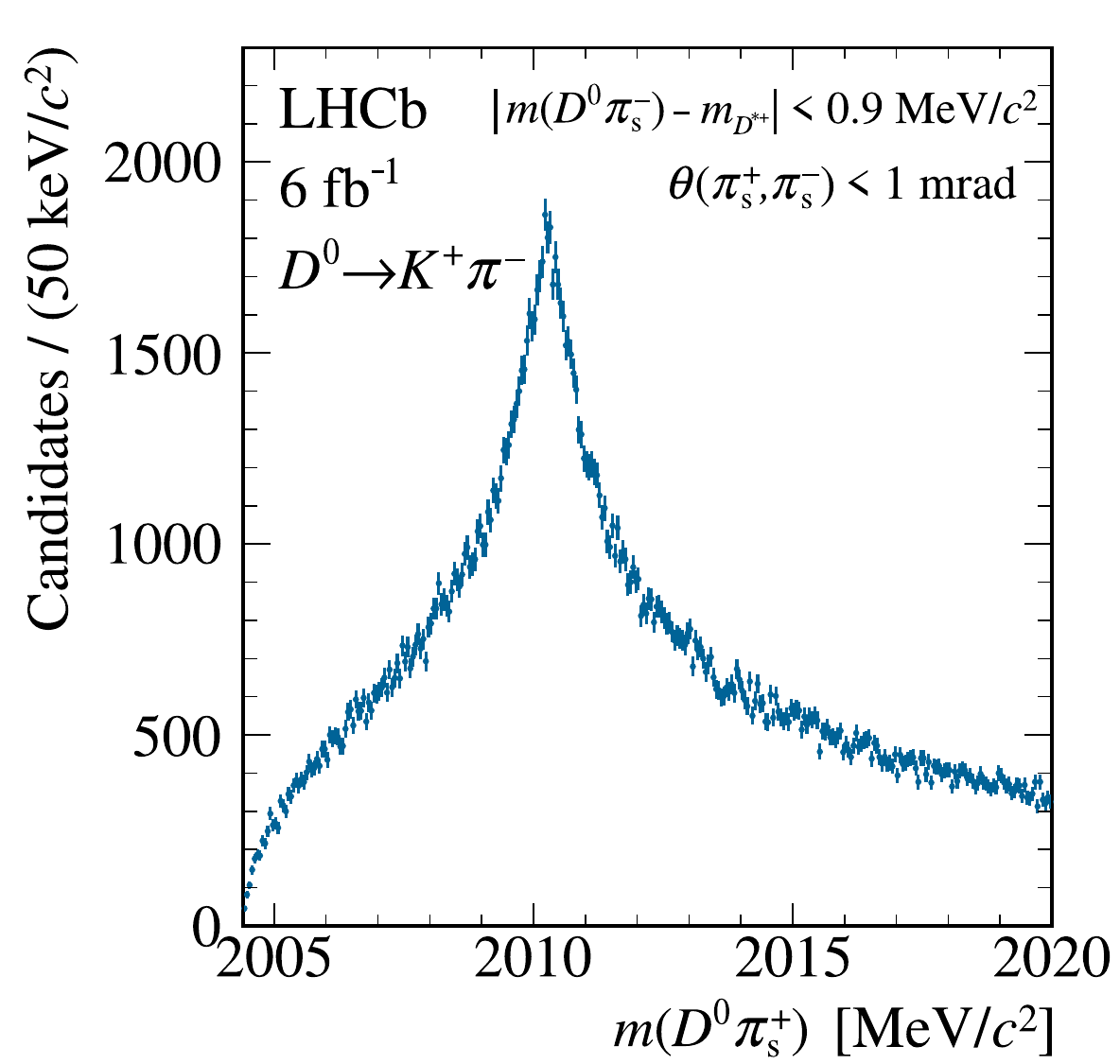}
		\caption{(Left) Distribution of the angle between common WS and RS soft pions, $\theta(\pi_\text{s}^{+},\pi_\text{s}^{-})$, for the common candidates subsample. The $1\mrad$ threshold, utilized to select the common ghost candidates, is marked by a vertical dashed line.
        (Right) Distribution of \mDp for the common ghost candidates.
  \label{fig:common_ghost}}
\end{center}
\end{figure}

As described in \cref{sec:selection}, whenever a \Dz candidate is used to reconstruct both a WS \Dstarm and a RS \Dstarp candidate, and the \mDp value of the RS candidate lies in the vicinity of the known \Dstarp mass~\cite{PDG2022}, the WS candidate is discarded.
The RS candidates belonging to this sample are mostly genuine \Dstarp signal decays, while the corresponding WS candidates arise from the association of the \Dz meson with a ghost soft pion or with an uncorrelated particle (in most cases, a pion originating from the PV). 
The opening angle between the directions of the soft pions of the WS and RS candidates, $\theta(\pi_\text{s}^{+},\pi_\text{s}^{-})$,  allows these two different sources of backgrounds to be disentangled, as shown in the left panel of \cref{fig:common_ghost}.
The distribution of the pairs where one of the two soft pions is a ghost has a narrow peak close to zero since they share the same clusters of hits in the vertex detector.
The component due to the uncorrelated particles, instead, has a much wider distribution, populating higher angle values.
The requirement $\theta(\pi_\text{s}^{+},\pi_\text{s}^{-}) < 1\mrad$ selects a pure sample of WS candidates associated with a ghost soft pion, referred to in the following as \textit{Common Ghost} (CG) candidates, and their \mDp distribution is shown in the right panel of \cref{fig:common_ghost}. 
The CG candidates are used in the analysis as a proxy for the residual ghost candidates present in the final WS (and RS) data sample. The CG candidates and the residual ghost candidates share the same features, except that they lie in two different kinematic regions of the reconstructed soft pion.
The CG candidates populate regions where both the genuine RS candidate and the ghost WS candidate are reconstructed within the acceptance of the tracking stations located downstream of the magnet.
The contamination of residual ghost soft pions in the signal sample, however, is mainly spatially distributed at the borders of the geometric acceptance, populating regions of the soft-pion kinematics with high charge asymmetries where, while the genuine associated RS candidate is not reconstructed, the ghost WS candidate is.
Hence, this residual background of ghost candidates is highly reduced by the fiducial requirement of \cref{eq:fiducial}, which removes kinematic regions with large charge asymmetries in reconstructing low-momentum particles.

The features of both CG candidates and the residual ghost candidates present under the WS (and RS) signal peaks are approximately reproduced by employing a data-driven technique~\cite{phd-thesis-roberto}.
By selecting a pair of genuine RS candidates in data with reconstructed clusters of hits in the vertex detector very close in space, an artificial sample of ghost candidates is generated by recomputing the \mDp invariant mass of one of the two RS candidates using the momentum magnitude of the soft pion of the other one.
This is repeated both for a pair of RS \Dstarp candidates within the acceptance of the final selection, passing the fiducial requirement of \cref{eq:fiducial}, and for a second pair of genuine RS candidates where only one candidate falls in the fiducial region.
The distributions of the \mDp invariant mass for the two artificial samples are found to be compatible, validating the usage of the CG candidates as a proxy for the residual ghost background in the sample. Compatibility is also assessed between the \mDp distribution of these artificial samples and that of the CG sample, apart from a small smearing that is needed because the directions of the two-track segments in the vertex detector do not exactly coincide. 

A complementary requirement on the angle,  $\theta(\pi_\text{s}^{+},\pi_\text{s}^{-}) > 1\mrad$,  selects a high-purity sample of WS candidates associated with unrelated particles. 
This is a sample of pure combinatorial background and is used to check the reliability of the empirical function used to model this type of background component.

\section{Ratio and average decay-time determination}
\label{sec:mass_fit}

The raw WS-to-RS yield ratios, $r^+_i$ and $r_i^-$ for the $\Kp\pim$ and $\Km\pip$ final states, respectively, and for each subsample $i$, are determined via a simultaneous \chisq fit to the \Dstarp invariant mass, \mDp, of WS, RS and CG candidates using empirical PDFs for the signal, combinatorial and ghost background shapes. While the distribution of WS and RS candidates receives contributions from these three components, that of CG candidates is only described by the ghost background shape. All parameters describing the PDFs are determined independently in each time interval from the mass fit, unless explicitly stated otherwise.

The PDFs of both WS and RS signal are modeled as the product of two functions, which together describe the distribution of the sum of promptly produced and secondary \Dstarp decays. 
The first one is the convolution of a Lorentz function with the sum of two Johnson $S_U$ functions~\cite{Johnson:1949zj}, while the second function serves to enforce the kinematical threshold of the \Dstarp decay, and it is equal to 
\begin{equation}
    \theta(\mDp-m_0) \cdot (\mDp-m_0)^{\rho}\,,
\end{equation}
where $\theta$ is the Heaviside step function and $m_0$ and $\rho$ are free parameters. The parameter $m_0$ is approximately equal to the sum of the \Dz and the \pip masses.
The signal shape is constrained to be the same for the WS and RS candidates in each bin, except for a global shift to the mean that takes into account small inaccuracies in the calibration of the momenta of particles of opposite charge. 

The PDF of the combinatorial background, both for WS and RS candidates, is proportional to  
\begin{equation}
\sqrt{\mDp-m_0}\times \left[1 + \alpha(\mDp-m_0)+\beta(\mDp-m_0)^{2}\right]\,,
\label{eq:pdf_combinatorial}
\end{equation}
where $\alpha$ and $\beta$ are small free parameters accounting for the deviation from a pure square-root behavior, and the $m_0$ parameter is shared with that of the WS and RS signal mass shapes. The PDFs of WS and RS combinatorial backgrounds have the same functional shape, but employ two different sets of parameters to account for very small contamination from misreconstructed decays of charmed mesons that may differ between RS and WS samples, but display a similar square-root behavior.

The ghost background shape is empirical and is described by the weighted sum of a Johnson $S_U$ function and a uniform distribution, multiplied by a term that enforces the kinematic threshold, as done for the WS and RS signal PDFs. 
Data-driven studies of this background guarantee the reliability of some simplifying assumptions~\cite{phd-thesis-roberto}, which are made to ensure the convergence of the fits. 
The shape parameters of this component are required to be the same for the WS, RS and CG candidates, and most of them are fixed to the result of a fit to the time-integrated sample. 
The shape of the ghost background candidates is found to depend only weakly on the \Dz decay time, as precisely verified by comparing the distribution of CG candidates among different decay-time intervals. Moreover, the absolute number of ghost candidates is constrained to be equal in WS and RS candidates, since they are mainly generated by the same random source. This latter assumption is checked by repeating the mass fits with and without the constraint and no significant change in the final results is observed. This is expected because the fraction of ghost soft pions in the RS sample is negligible; however, this constraint is kept to improve the stability and the reproducibility of the mass fits.

\begin{figure}
    \centering
    \includegraphics[width=0.485\linewidth]{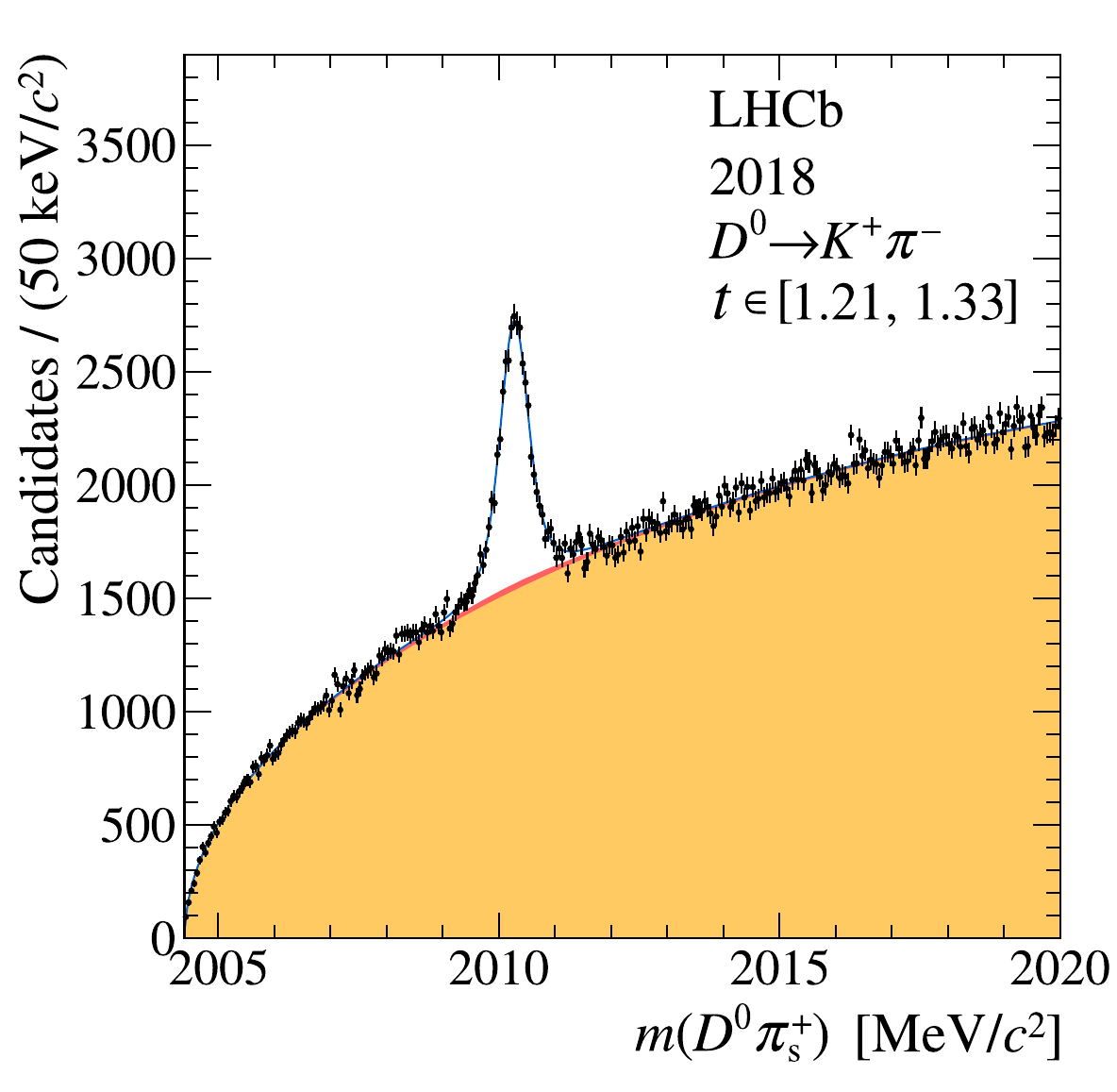}
    \includegraphics[width=0.485\linewidth]{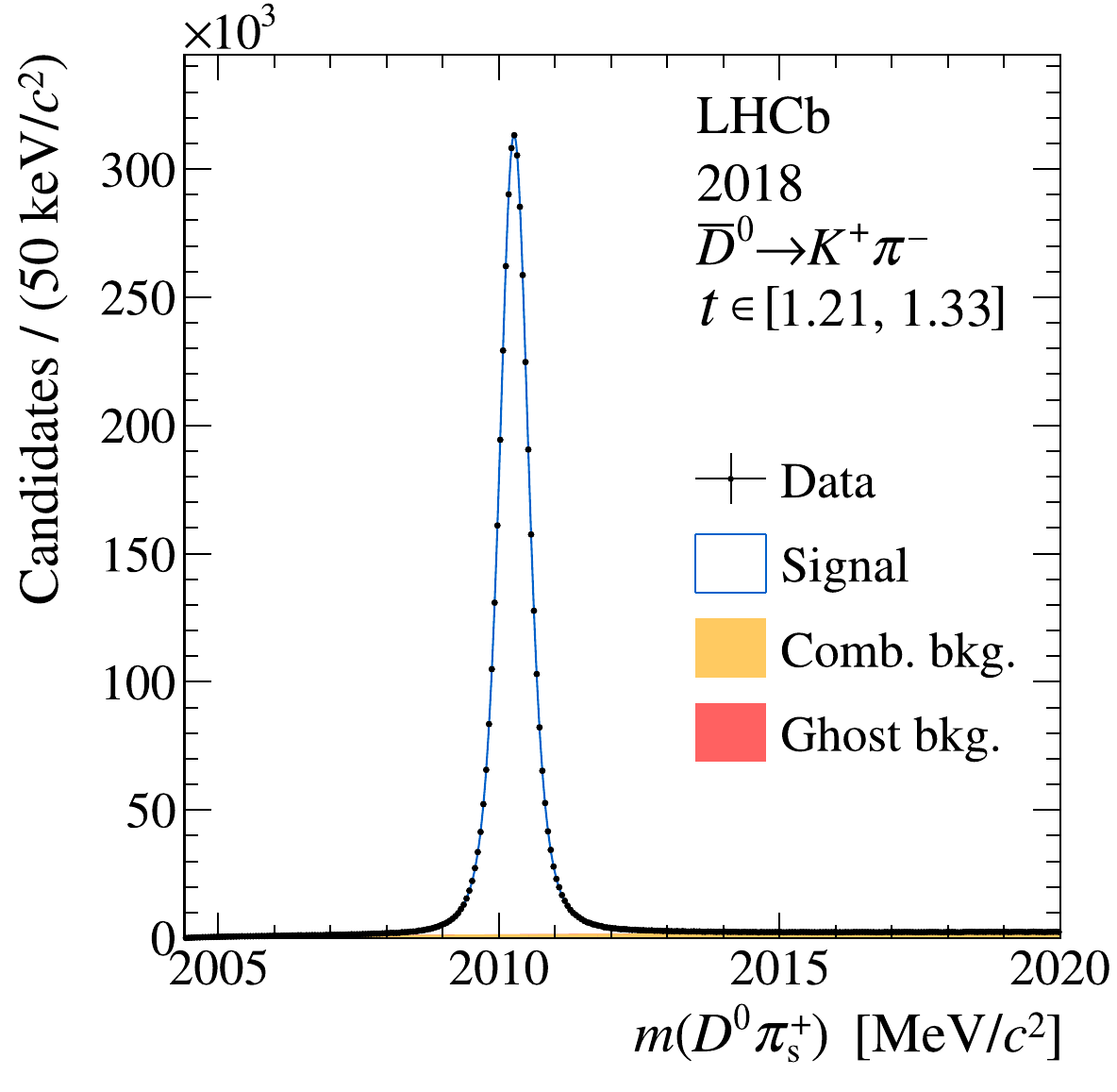}\\
    \includegraphics[width=0.485\linewidth]{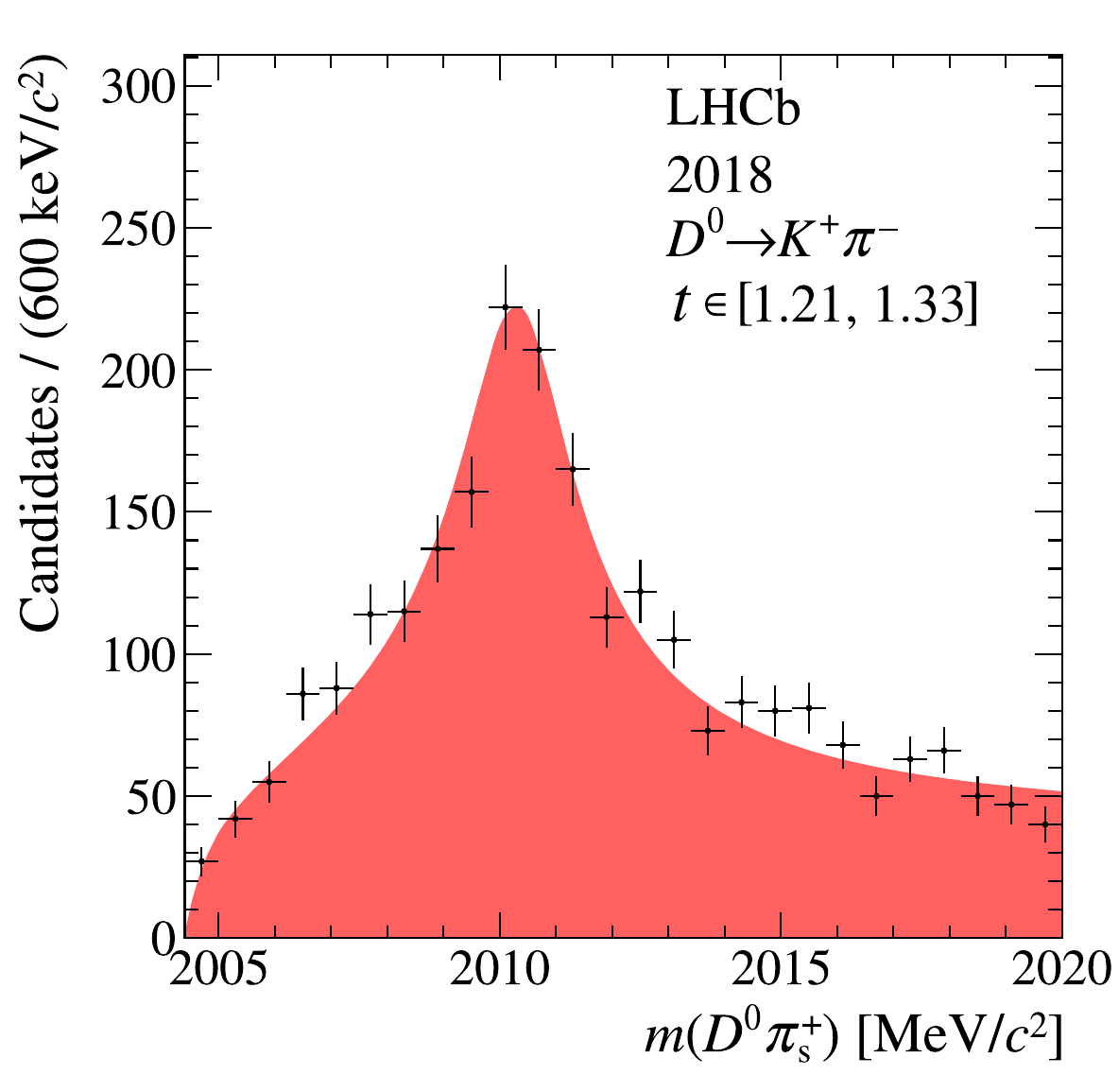}
    \caption{An example of the \mDp distributions of (left) WS, (right) RS, (bottom) and CG events of the 2018 $\Kp\pim$ sample in the $t\in[1.21,1.33]$ decay-time interval. The value of the \chisq per number of degrees of freedom of this fit is 653/631. Different fit components are superimposed as indicated in the legend.}
    \label{fig:mass_fit}
\end{figure}

\Cref{fig:mass_fit} shows the projections of a typical simultaneous fit, in a given decay-time interval, to the \mDp invariant-mass distribution of the RS, WS, and CG candidates. 
For RS and WS decays, different contributions from signal, combinatorial background, and CG decays are also displayed. 
%%%%
The distribution of the measured values of the $\chi^2$ per number of degrees of freedom (\chisqndf) of all 108 mass fits is found to display a Gaussian-like shape to a good approximation. The mean value is about 1.06, while its standard deviation equals about 0.07, to be compared with the expected values of 1 and $0.06$, respectively.
An inflation factor of $\sqrt{1.06}$ is applied to the measured uncertainties of the WS-to-RS signal ratios to account for possible small effects of the mismodeling of the empirical PDFs used.
Furthermore, many checks and studies are performed on both the reliability and robustness of the adopted strategy for the fit to the \Dstarp invariant-mass distribution~\cite{phd-thesis-roberto}.

The average values of the decay time and of the squared decay time, $\langle t \rangle_i$ and $\langle t^2 \rangle_i$, respectively, are evaluated in each subsample $i$ using data after removing the contribution of the combinatorial background by means of a sideband-subtraction procedure in the \mDp distribution. Background candidates in a lateral mass window, $\mDp \in [2014,2020]\mevcc$, are weighted with a suitable negative coefficient, while  candidates in the the signal region, $\mDp \in [2009.37,2011.17]\mevcc$, are left unchanged with weights equal to unity. The negative coefficient is defined as the ratio of the integral over the signal region to that over the lateral mass window of the combinatorial background PDF. The analytical model of this PDF is is determined by a mass fit similar to the one described above. Here, the constraint that the \Dstarp candidate originates from the \PV is not used in the vertex fit, to include all secondary decays, which otherwise would be artificially migrated to lower masses. The ghost background component is negligible and therefore not taken into account in the fit. Uncertainties on $\langle t \rangle_i$ and $\langle t^2 \rangle_i$ are very small compared with other uncertainties on decay-time biases and are therefore neglected. 

\section{Sources of biases}
\subsection{Ratio bias}
\label{sec:ratio_bias}
Measurements described in Refs.~\cite{LHCb-PAPER-2017-046} and \cite{LHCb-PAPER-2020-045}, using a dataset largely overlapping with that used in this analysis, precisely quantified the contamination from singly misidentified two-body decays (\decay{\Dz}{\Kp\Km} and \decay{\Dz}{\pip\pim}) and misreconstructed multibody charm decays, in both RS and WS data samples. From these studies, it can be concluded that these backgrounds are negligible. 
The main background to the RS sample is given by \decay{\Dz}{\Km\ell^{+}\nu_{\ell}} decays, at a level of 0.03\% of the RS signal yield. The main background to the WS sample is given instead by \decay{\Dz}{\pip\pim\pi^0} decays at a level of about 0.1\% of the WS signal yield. However, this contribution has a wider distribution in \mDp, so that its contamination to the signal is further reduced when the fit to the invariant mass is performed. No evidence of \CP violation has been found in dedicated studies of this decay~\cite{LHCb-PAPER-2023-005, LHCb-PAPER-2024-003,BaBar:2008xzl}, and it can be safely neglected.  

\begin{figure}
    \centering
    \includegraphics[width=0.6\linewidth]{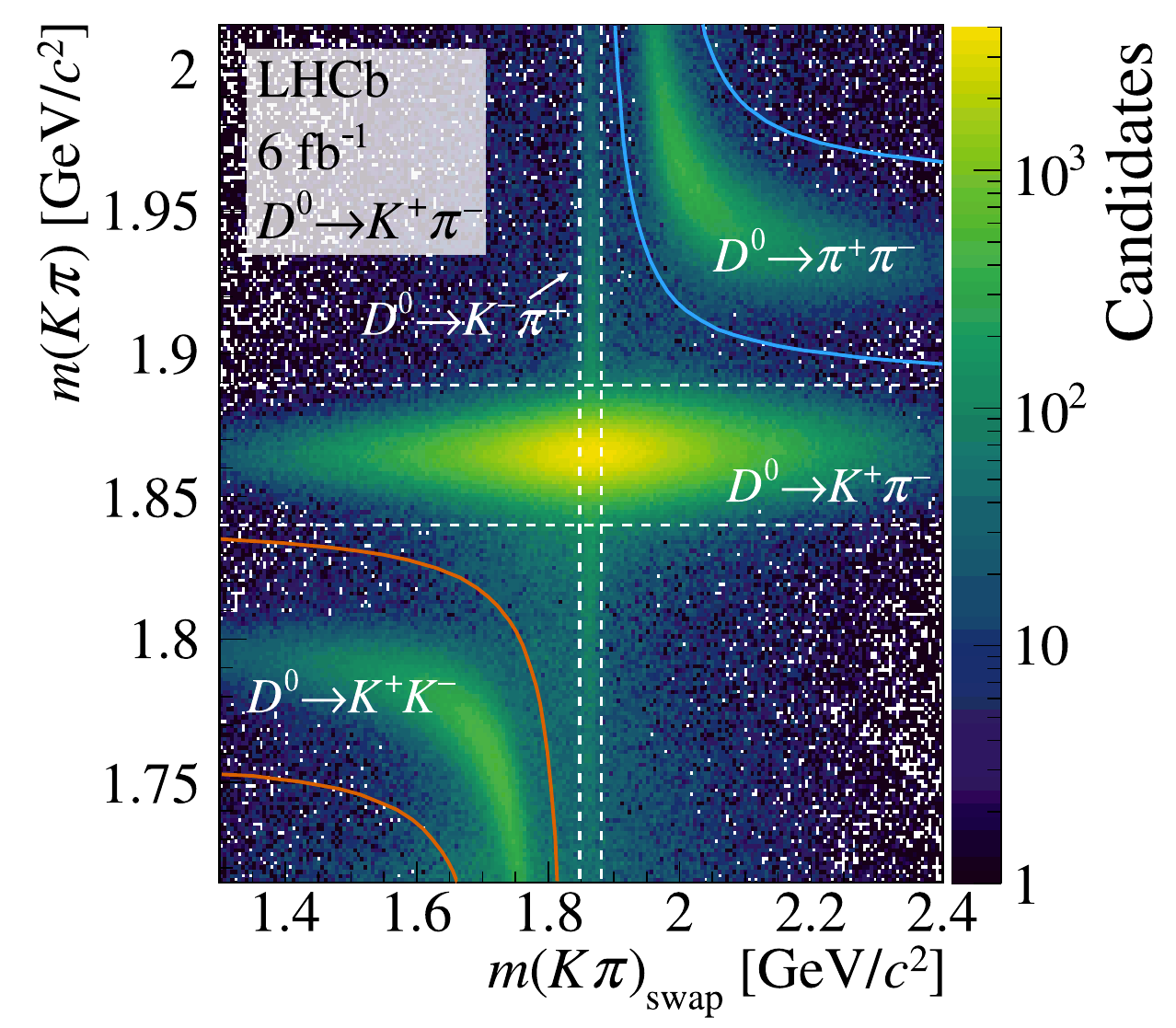}
    \caption{Background-subtracted two-dimensional distribution of the invariant \kaon\pion mass, computed with the standard mass hypothesis, $m(\kaon\pion)$, \vs  the swapped mass hypothesis, $m(\kaon\pion)_{\textrm{swap}}$, for WS candidates. Horizontal (vertical) dashed lines indicate the signal (vetoed) region, while blue (orange) solid lines indicate the region where the reconstructed invariant \Dz mass is within $\pm 40\mevcc$ of its known value when computed with the \pion\pion (\kaon\kaon) mass hypothesis.  }
    \label{fig:Kpi_vs_piK}
\end{figure}

The residual contamination from doubly misidentified \decay{\Dz}{\Km\pip} decays can mimic a time-dependent effect since it is correlated with the momentum of the \Dz meson. It is estimated via a fit to the two-dimensional distribution of the \kaon\pion invariant mass computed with the correct and the inverted \kaon-\pion mass hypotheses, $m(\kaon\pion)$ \vs $m(\kaon\pion)_{\textrm{swap}}$, of background-subtracted WS candidates, where any requirement on the \Dz mass is removed, as shown in \cref{fig:Kpi_vs_piK}. After applying the requirements on both \Dz masses, the bias to the measured ratio is estimated to be $2\times10^{-6}$ and is subtracted (see Sec.~\ref{sec:mix_fit}). A conservative uncertainty equal to half the bias value is assigned. Time-dependent variations of this bias are smaller and are, therefore, neglected.
\begin{figure}
    \centering
\includegraphics[width=0.6\linewidth]{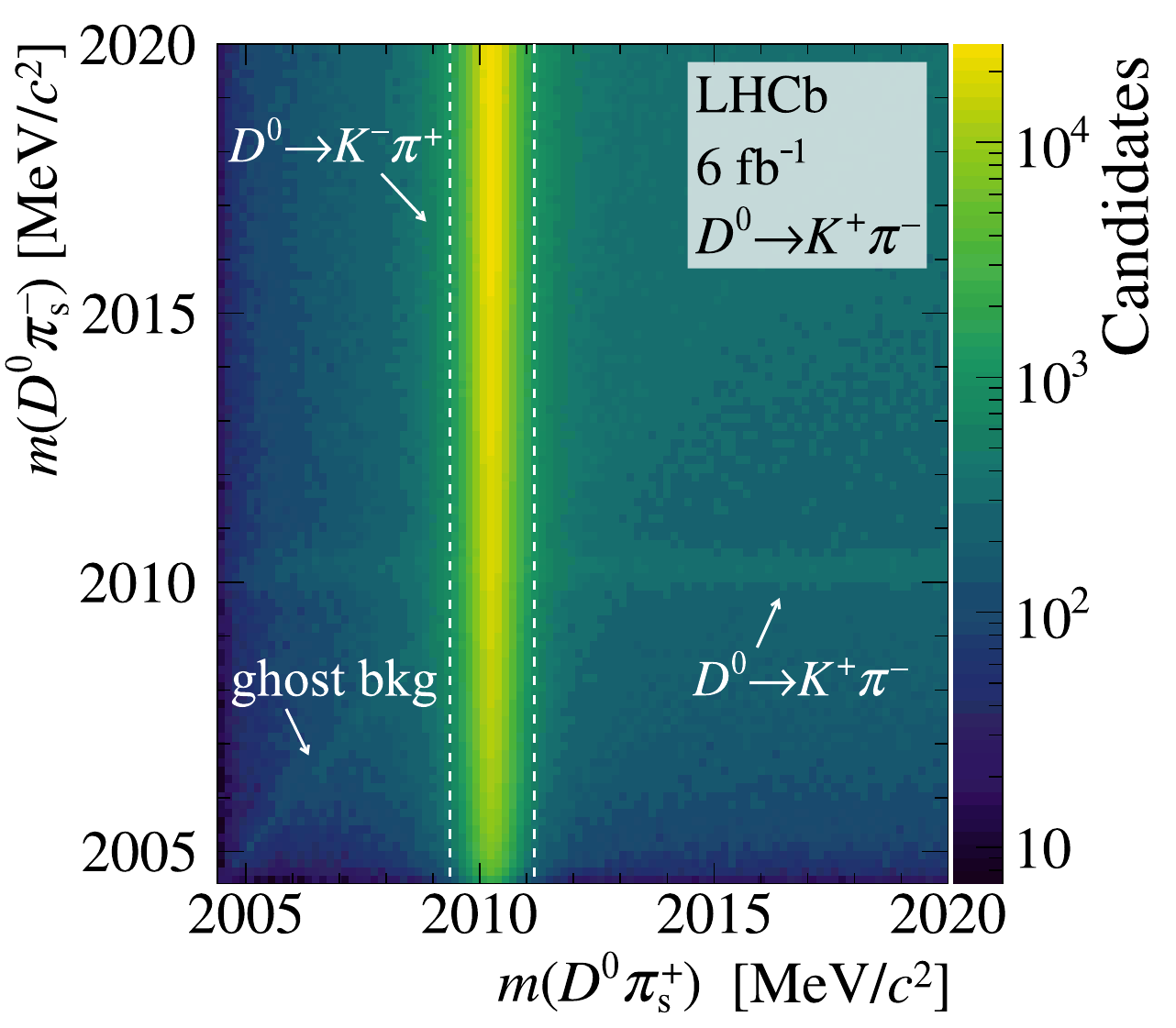}
\caption{Two-dimensional distribution of the invariant masses of the \Dstarp and \Dstarm mesons reconstructed from the same \Dz meson, but using two different candidates for the soft pion. Candidates in the region within the vertical dashed lines are discarded from the WS sample, as described in \cref{sec:selection}. The barely visible diagonal band is due to ghost soft pions.
}
    \label{fig:common_WS_vs_RS}
\end{figure}

The fraction of signal WS decays removed by vetoing the RS and WS common candidates is determined by fitting the two-dimensional distribution of the invariant masses of the common \Dstarp and \Dstarm candidates reconstructed from the same \Dz meson, but with oppositely charged soft pions, which is shown in \cref{fig:common_WS_vs_RS}. The measured fraction is about $0.13\%$ and the corresponding bias is subtracted (see Sec.~\ref{sec:mix_fit}). A conservative uncertainty equal to half the bias value is again assigned, and the time-dependent variations are neglected.

\subsection{Asymmetry bias}
\label{sec:asymmetry_bias}
Nuisance charge asymmetries mainly originate from the different probabilities of producing \Dstarp and \Dstarm mesons in a $pp$ collision and from the different efficiencies of detecting positively and negatively charged low-momentum particles, such as the soft pion utilized to form the \Dstarp candidates. 
The \lhcb detector is not perfectly left-right symmetric, and the detection and reconstruction process, including pattern recognition, track reconstruction, and selections, is intrinsically charge-asymmetric.
The net effect of such instrumental asymmetries is a bias to the WS-to-RS yield ratio, acting in opposite directions for the $\Kp\pim$ and $\Km\pip$ final states, and mimicking a \CP-violating asymmetry indistinguishable from a real effect due to the effective Hamiltonian for \Dz mesons.  Any such asymmetry must be precisely removed.

For a given subsample~$i$, the measured biased value of the WS-to-RS yield ratio, \tRipm, is related to the unbiased value, \Ripm, through the following expression 
\begin{equation}\label{eq:asym_corr}
\tRipm = \Ripm\,\left(1\pm 2\AIi\right),
\end{equation}
where \AIi is the instrumental asymmetry. This asymmetry is measured using a pure calibration sample of 40 million \decay{\Dstarp}{\Dz(\to\Kp\Km)\spip} decays, collected in the same conditions and with almost identical requirements as the RS and WS data samples. The raw asymmetry, \takki, is measured from data in each subsample $i$ by counting the number of reconstructed \decay{\Dz}{\Kp\Km} (\decay{\Dzb}{\Kp\Km}) candidates, $N_i^+$ ($N_i^-$), as 
$\takki = (N_i^+ - N_{i}^-)/ (N^+_i + N_{i}^-)$.
The instrumental asymmetry is determined as
\begin{equation}
\AIi  = \takki - (\adkk  + \DY\,\langle t \rangle_i)\,.
\label{eq:instr_asym}
\end{equation}
The terms \adkk and $\Delta Y$ are the \CP-violating asymmetry in the decay and the time-dependent \CP asymmetry in the singly Cabibbo-suppressed \decay{\Dz}{K^+K^-} mode, respectively. They are both external inputs to this analysis and their measured values~\cite{LHCb-PAPER-2022-024,LHCb-PAPER-2020-045}, with the associated uncertainties, are used as nuisance parameters in the fit to the time-dependent WS-to-RS yield ratios. The average value of the decay time, in each sub-sample, is approximately equal to that of RS candidates, and is denoted by $\langle t \rangle_i$. The derivation of \cref{eq:asym_corr,eq:instr_asym}, 
reported in Ref.~\cite{phd-thesis-roberto}, 
assumes small values of both production and detection asymmetries, including those of the soft pion, over the whole kinematic domain. This is enforced by the fiducial requirement on the soft-pion momentum given in \cref{eq:fiducial}, which allows also neglecting higher-order terms proportional to the detection asymmetry of the $\kaon\pion$ pair.

\begin{figure}
    \centering
    \includegraphics[width=0.48\linewidth]{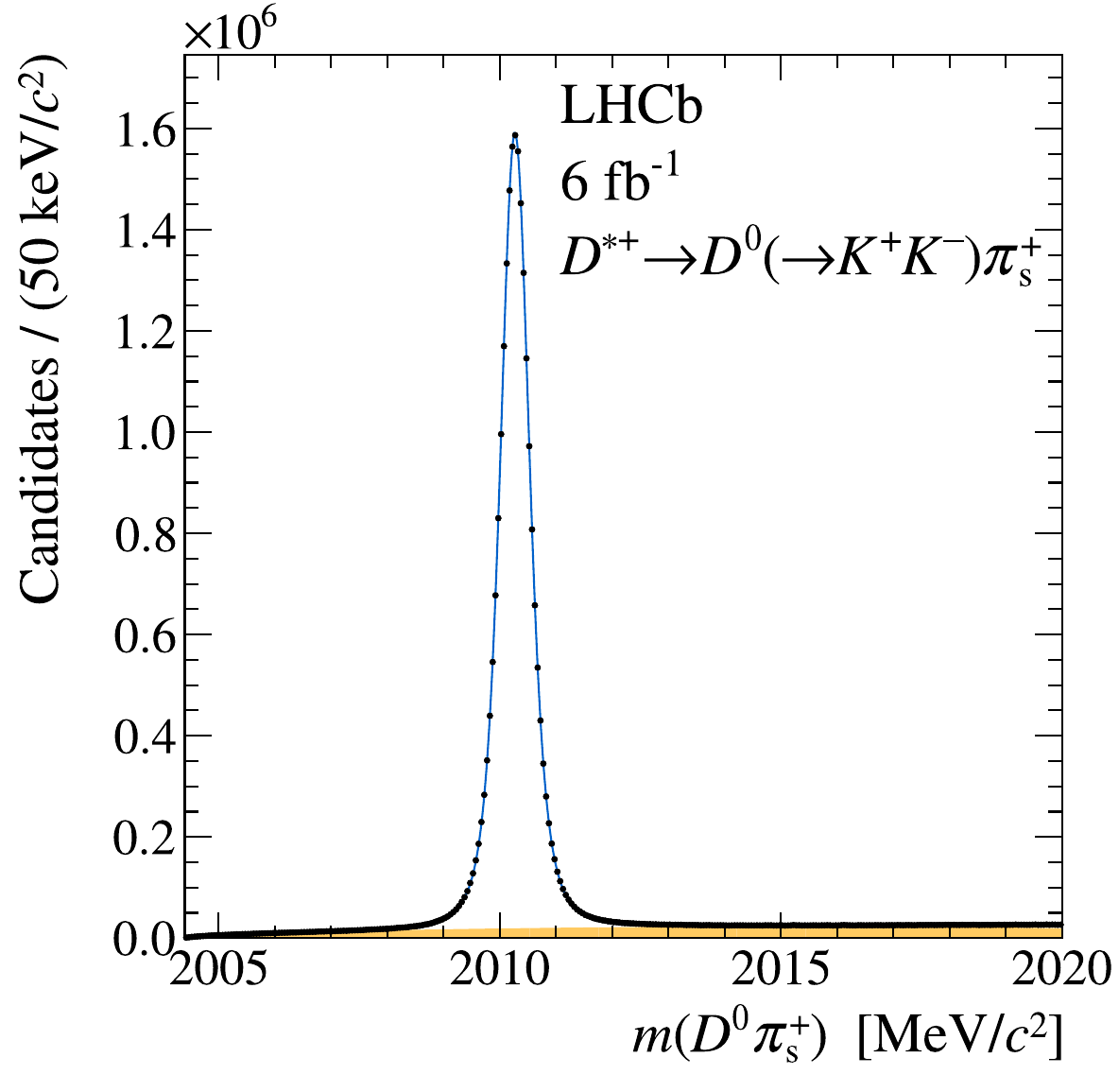}
    \includegraphics[width=0.48\linewidth]{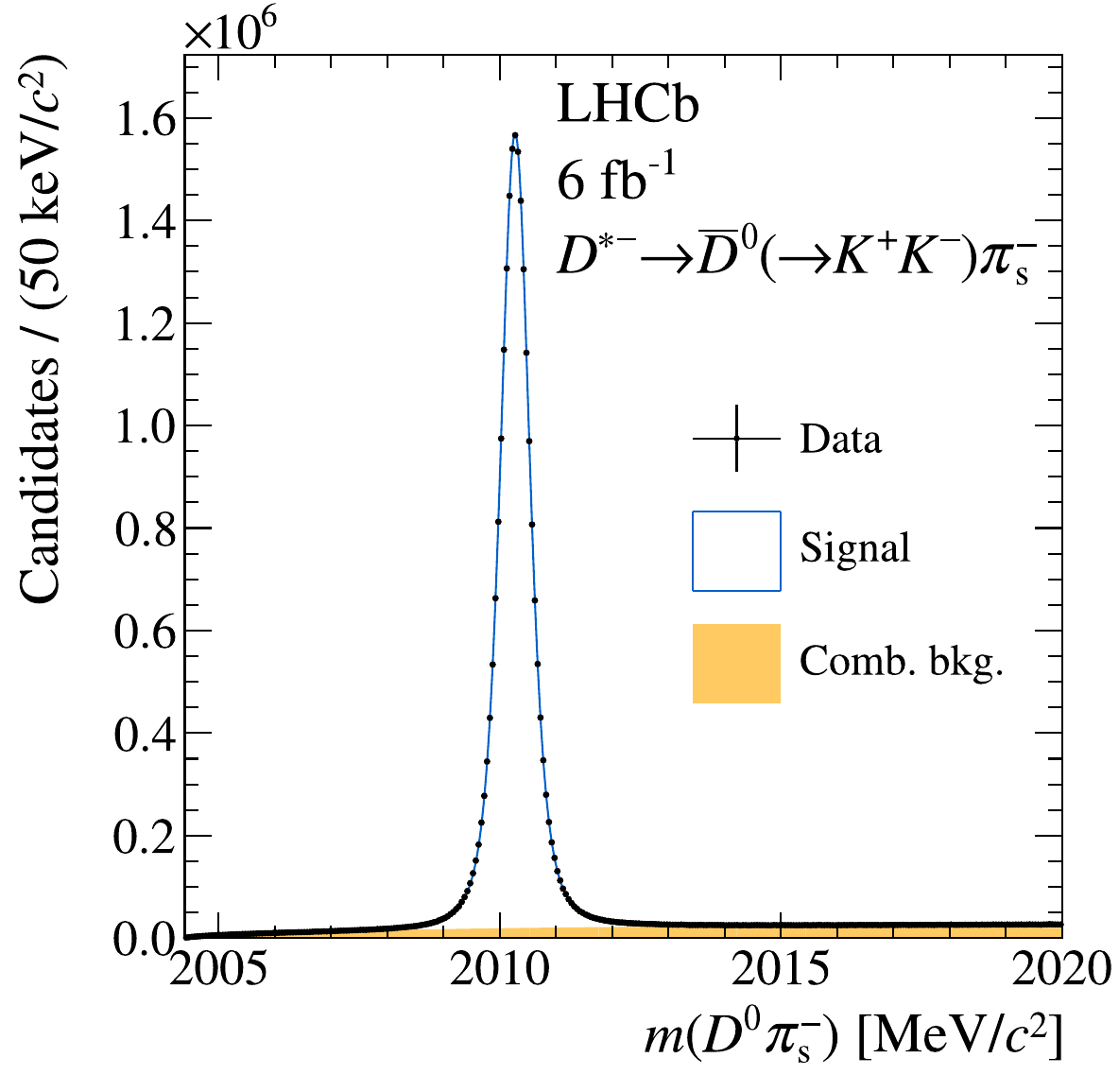}
    \caption{Mass distributions of (left) \decay{\Dz}{K^+K^-} and (right) \decay{\Dzb}{K^+K^-} candidates after the offline selection, with results of the fit described in the text superimposed.}
    \label{fig:final_sample_KK}
\end{figure}

\begin{figure}
    \centering
    \includegraphics[width=0.55\linewidth]{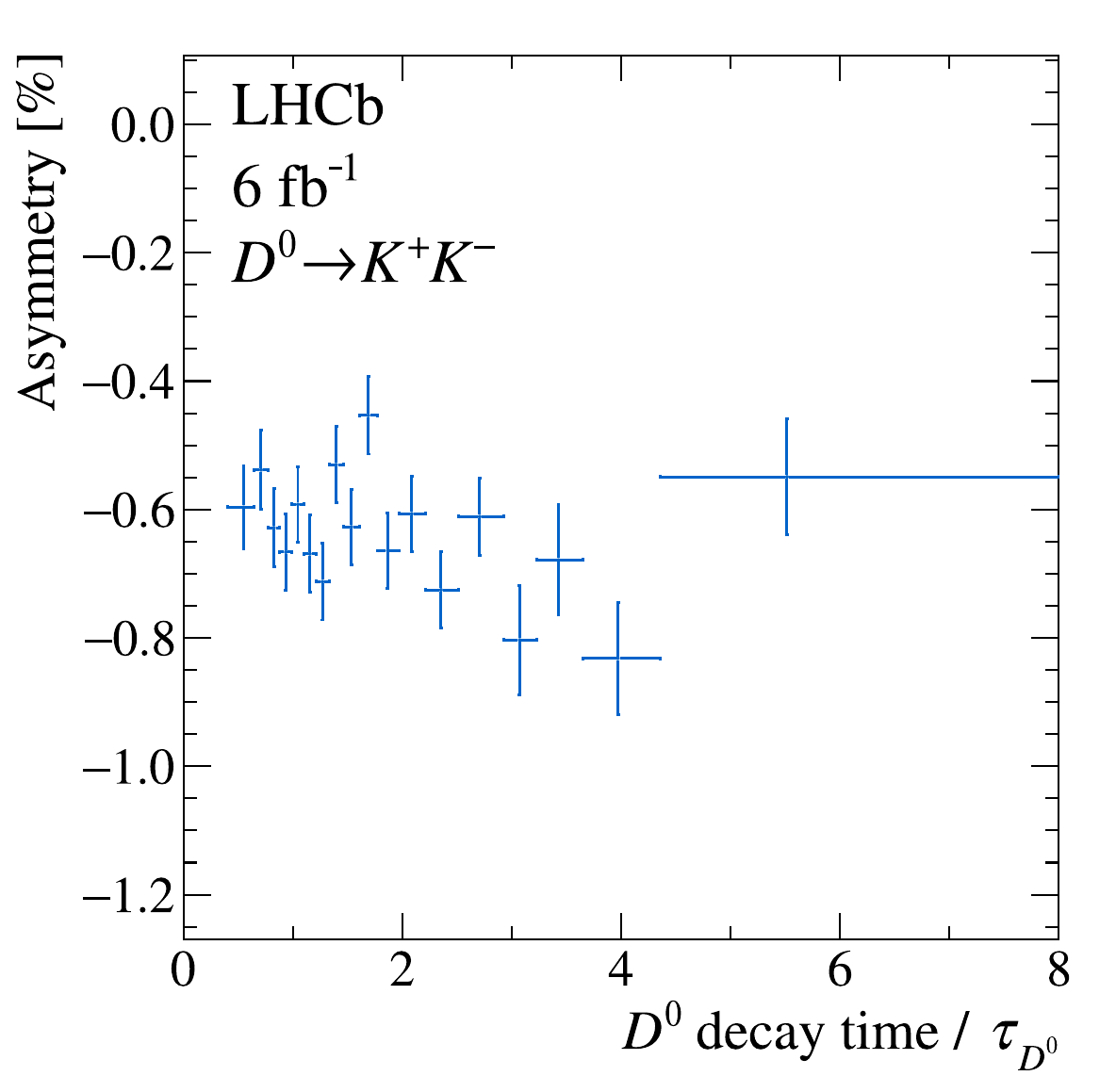}
    \caption{Measured values of the raw asymmetry, \takki, in each decay-time interval, for the \decay{\Dz}{\Kp\Km} decays. Results are averaged over the different data-taking periods.}
    \label{fig:instr_asymmetry}
\end{figure}

The numbers of \Dz and \Dzb to $\Kp\Km$ decays in each subsample are determined by means of a simultaneous binned $\chi^2$ fit to the \mDp invariant-mass distribution of the \Dstarp and \Dstarm candidates, which employs the same PDFs as in \cref{sec:mass_fit} for the signal and the combinatorial background, but neglects the ghost background.
The \mDp distributions of \Dz and \Dzb candidates after all selections are shown in \cref{fig:final_sample_KK}, with the result of a fit superimposed.
The quality of the fits is inspected in each subsample and no sign of mismodeling is found.
%%%
The raw asymmetries are found to be stable across all data-taking periods and do not show significant decay-time dependency, as shown in \cref{fig:instr_asymmetry}.

In order to measure instrumental asymmetries, \AIi, in identical conditions as for RS and WS decays, a weighting procedure is applied, separately for each data-taking period and \Dz decay-time interval, to equalize the kinematics of \decay{\Dz}{\Kp\Km} decays to that of \decay{\Dz}{\Km\pip} decays. Weights are computed by comparing the six-dimensional ($\pt(\Dz)$, $\eta(\Dz)$, $\phi(\Dz)$, $\pt(\spip)$, $\eta(\spip)$, $\phi(\spip)$) background-subtracted distributions between the two samples, where $\phi$ is the azimuthal angle with respect to the $x$ axis. 
The weights are distributed close to unity since the kinematics of the two samples are very similar.

\subsection{Decay-time bias}
\label{sec:time_bias}
The contamination from secondary decays is the main source of bias to the \Dz decay-time determination.
Requirements on observables sensitive to decay-time resolution at trigger level and bin migration are minor sources of decay-time bias which also affect promptly produced \Dstarp decays, hereafter called prompt decays.
The total biases affecting  the measured values of $\langle t \rangle$ and $\langle t^2 \rangle$, in each subsample $i$, are equal to
\begin{align}
  \langle \delta t \rangle_i  =\,& \langle \delta t \rangle_i^\text{P}(1-f_i) + \langle \delta t \rangle_i^\text{S} f_i\,,\\
  \langle \delta t^2 \rangle_i =\,& \langle \delta t^2 \rangle_i^\text{P}(1-f_i) + \langle \delta t^2 \rangle_i^\text{S} f_i\,,
\end{align}
where the superscripts P and S indicate that the average is that for prompt and secondary decays, respectively, and $f_i$ is the fraction of secondary decays in each bin $i$.
The values of $\langle \delta t \rangle_i$  and $\langle \delta t^2 \rangle_i$ for prompt and secondary candidates and the relative fraction $f_i$, for each data-taking period are determined by performing a template fit to the two-dimensional $\IP(\Dz)$ \vs $t$ distribution of the background-subtracted RS candidates, where \Kp\pim and \Km\pip final states are combined. The discriminating power in disentangling the prompt and secondary component is increased by loosening the requirement on the \IP of the \Dz meson to $\IP(\Dz)< 600\mum$. 
This also allows checking that the distribution of the secondary decays, which mostly populate regions with high values of the \IP, is accurately modeled.

The two-dimensional templates are generated with a \geant-based simulation which simulates only prompt and secondary \Dstarp decays, neglecting all other particles emerging from the $pp$ collisions. 
The detector resolutions and efficiencies evaluated in these samples are expected to be superior to those of the full simulation due to the considerably lower detector occupancy.
However, this approach increases the speed of producing simulated events by approximately a factor of 50, and allows much larger samples of events to be saved on disk, with no trigger and offline requirements applied except those related to the \lhcb geometrical acceptance. Thus, the detector response can be accurately tuned to match data in relevant variables such as the vertex resolution. 
The absence of the underlying event emerging from $pp$ collisions causes loss of information on the reconstructed PV position.
This is addressed by smearing the true PV position of each simulated event according to the covariance matrix obtained from the PV fit in data, properly reproducing the behavior observed in data, including tails and dependence on track and vertex multiplicities.
To improve the level of agreement between data and simulation, the resolution of the \Dz decay vertex (DV) in simulation is weighted to be on average the same as that of data.  Moreover, a misalignment of about 10\mum is observed in data between the left and right halves of the vertex detector, which broadens the $\IP(\Dz)$ distributions. Since the size of this effect is not accurately reproduced in simulated samples, a misalignment of the same size is applied to match data.
Smaller simulated samples with the underlying event emerging from the $pp$ collisions are also available and are used to check all aspects of the analysis related to the full topology of the event, such as the probability of associating an incorrect PV to the \Dstarp candidate. 

The kinematics of the simulated \Dz mesons and soft pions are weighted in the multi-dimensional space of their momenta to align with data~\cite{Rogozhnikov:2016bdp}. The weights are computed using data sets with $\IP(\Dz) < 60\mum$ ($>120 \mum$) for prompt (secondary) decay candidates. The first subsample coincides with the signal sample, enriched with prompt decays. The contribution of the small residual contamination of secondaries, which is the target of this study, does not affect the momentum distributions and it is, therefore,  neglected for the purpose of weighing the simulated samples.  
The latter subsample is a pure sample of secondary decays.  

The sample of simulated secondary decays accounts for all known $b$ hadrons decaying into a generic final state with a \Dstarp meson. Production fractions of different hadrons and branching fractions of all the considered decay modes are taken from Ref.~\cite{PDG2022}. To account for the limited knowledge of the composition of this sample, a small number of \decay{\Bz}{\Dstarp X} decays, where $X$ is a particle of arbitrary mass, is simulated and added to the mixture. Both the relative fraction, $f_X$, of this decay mode and the mass, $m_X$, of the $X$ particle are treated as nuisance parameters in the template fit. This is intended to be an effective correction, as confirmed by the small size of observed discrepancies. In support of this approach, a satisfactory level of compatibility between data and simulation is found when the \Dstarp meson is required to form a good vertex with a charged muon. This is a very pure sample of $\decay{\Bzb}{\Dstarp \mu^-\bar{\nu}_{\mu}}$ decays, and it can be directly compared with simulated decays without any contamination from unknown decay modes and external inputs. 

\begin{figure}[tb]
    \centering
    \includegraphics[width=0.48\linewidth]{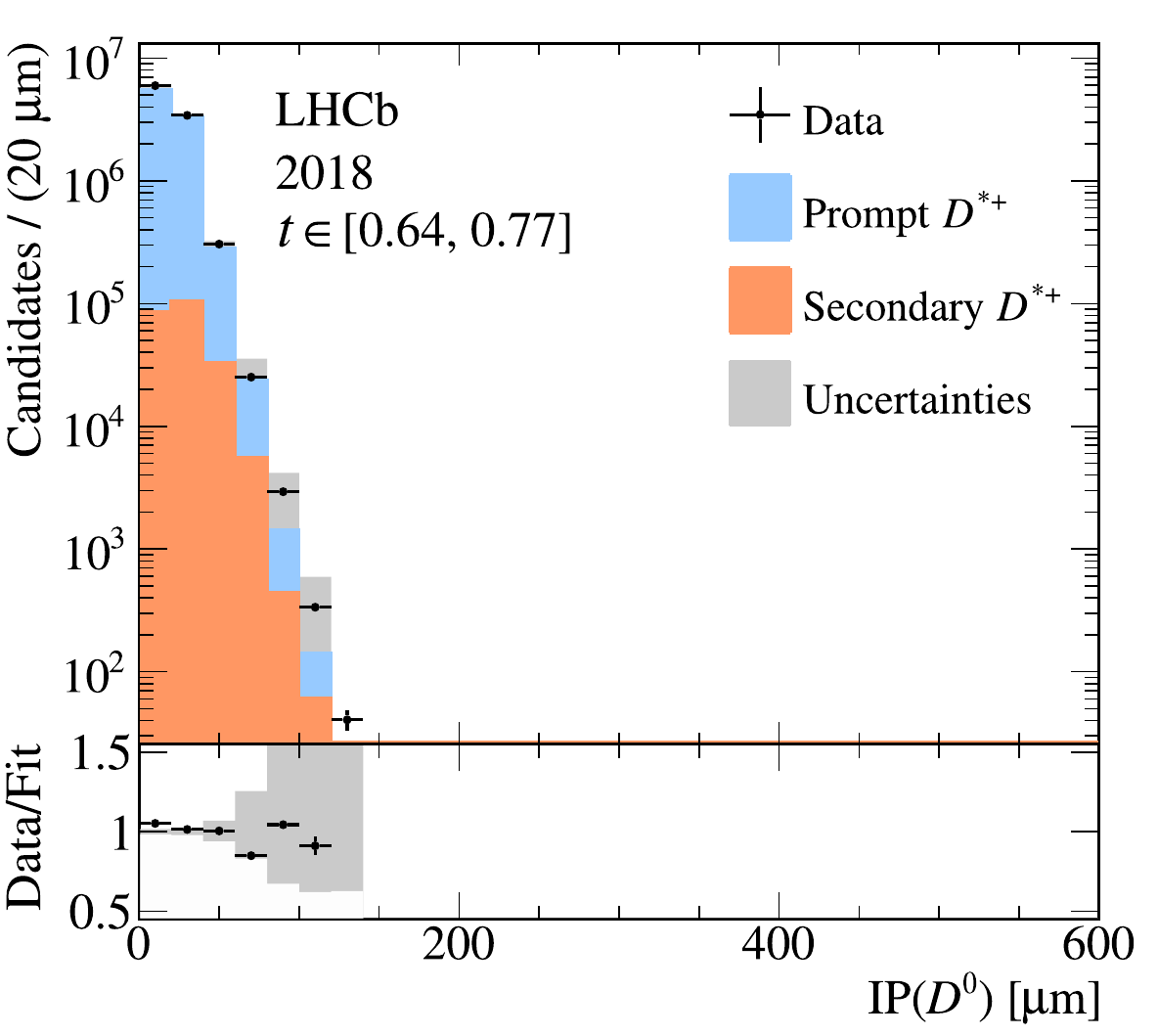}
    \includegraphics[width=0.48\linewidth]{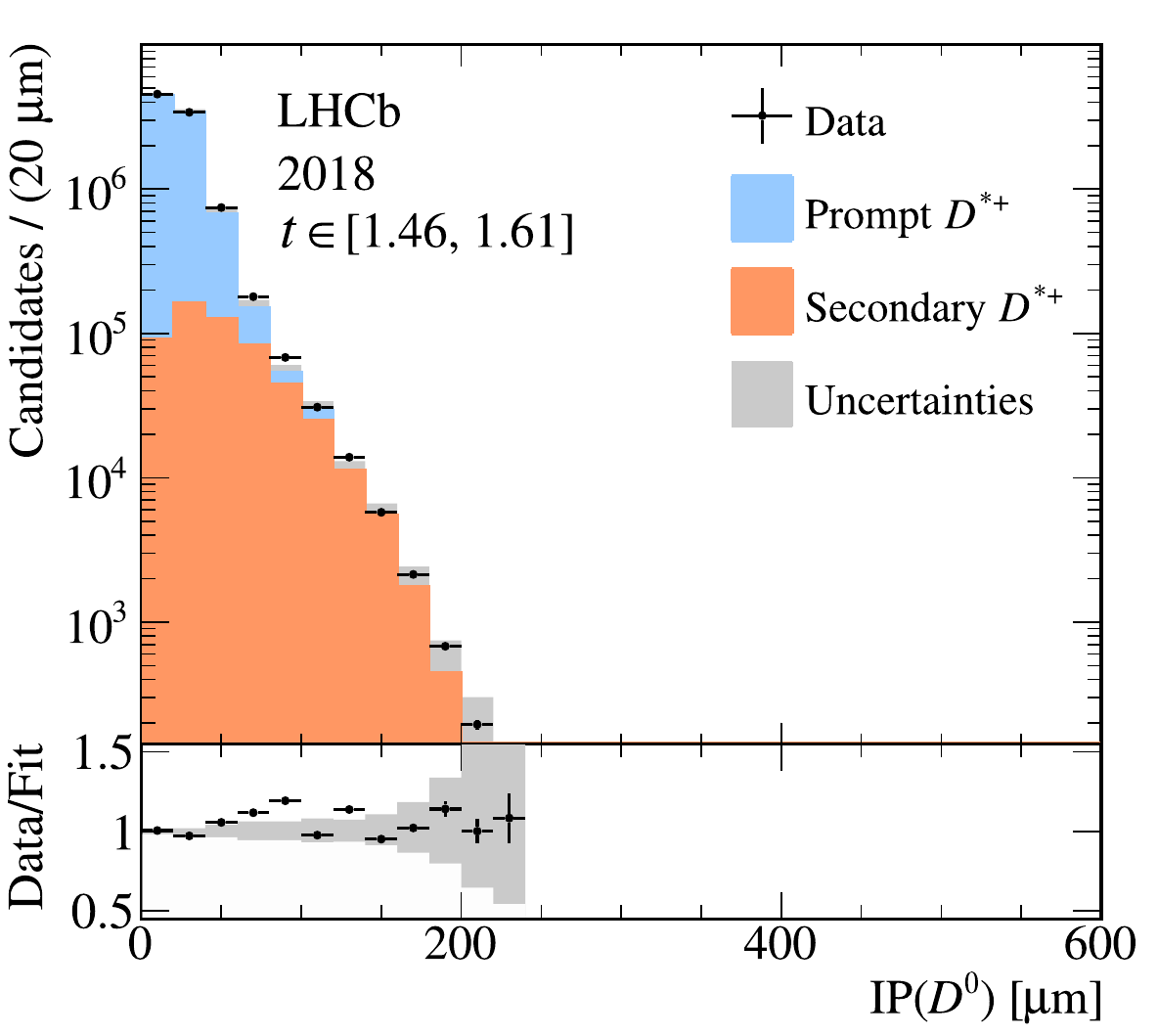}
    \includegraphics[width=0.48\linewidth]{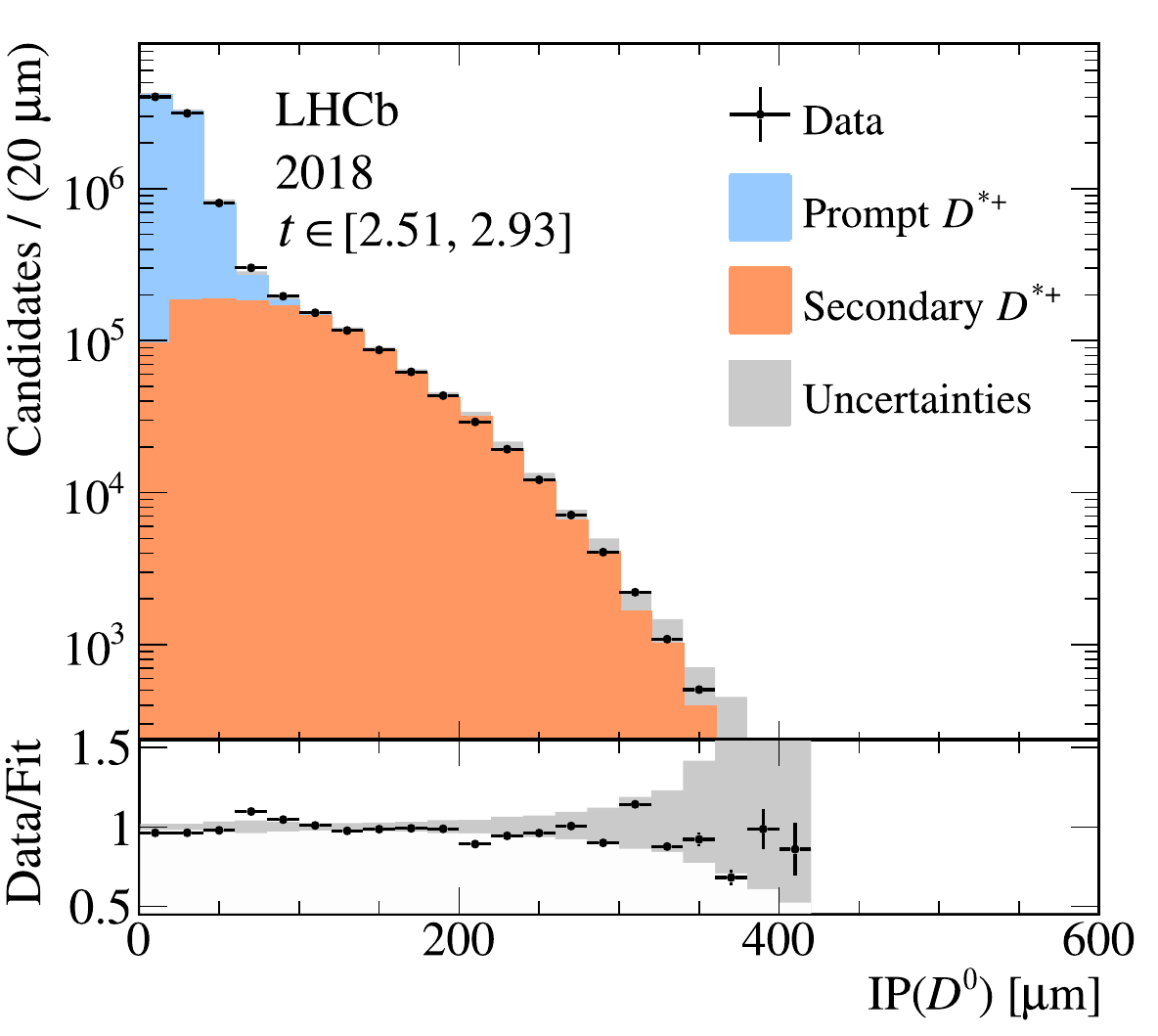}
    \includegraphics[width=0.48\linewidth]{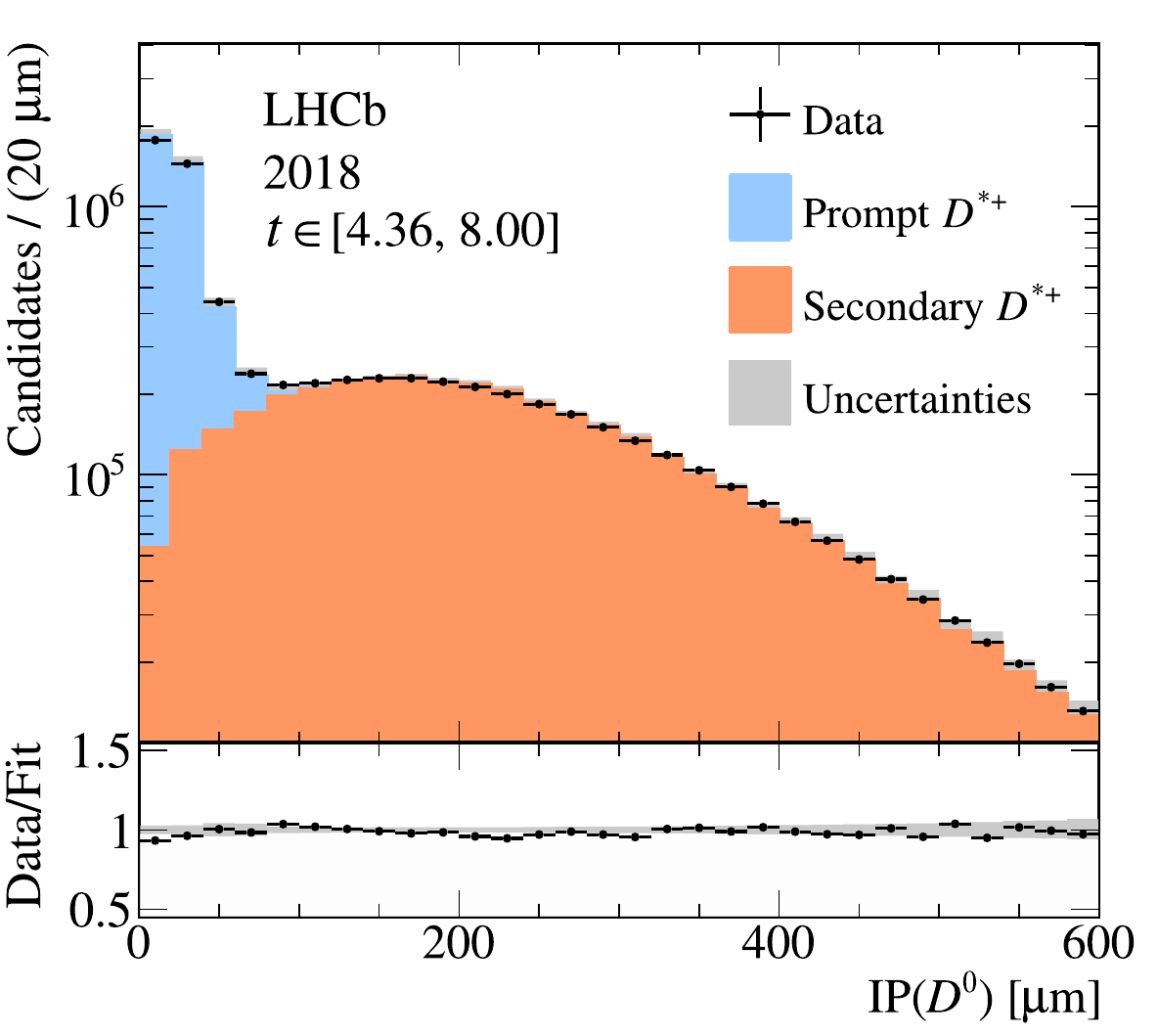}
\caption{Distributions of the IP of the RS sample candidates from the 2018 data-taking period for different decay-time intervals. The projections of the two-dimensional template fit
are superimposed. The points in the lower panel of each plot show the data-to-fit ratio, where the vertical black lines represent the statistical uncertainties of data, and the grey error band displays the total uncertainty of the simulation.}
    \label{fig:template_fit_projection}
\end{figure}

Systematic uncertainties from the limited knowledge of the PV and DV resolution, $f_X$ and $m_X$ are treated with the template profile likelihood approach~\cite{Cranmer:1456844}. Templates are produced with the PV and DV resolution independently scaled by a factor of 0.9, 1 and 1.1 of the baseline scenario, $f_X$ is chosen equal to 0\%, 3\% and 6\%, $m_X$ is chosen between 0.5, 1.5 and 2.5\gevcc.
Templates corresponding to intermediate values of these nuisance parameters are obtained through piece-wise linear interpolation.
Uncertainties on the number of events in the bins of the templates are treated using the Beeston--Barlow prescription\cite{Beeston-Barlow}.
The only free parameters in the fit relate to the normalization of the prompt and secondary templates, and the four nuisance parameters described above. The fitted values for the PV (DV) resolution scale factors are within 1.07--1.09 (1.02--1.04), while a relative fraction of about 1\% is found for an effective particle with mass $m_X \approx 0.8\text{\textendash}1.3\gevcc$.
Prompt and secondary simulated samples are then used to determine the biases $\langle \delta t^{(2)} \rangle_i^{P}$, $\langle \delta t^{(2)} \rangle_i^{S}$, and the relative fraction $f_i$ in each subsample $i$.
The agreement between data and fit projections is shown in \cref{fig:template_fit_projection}. The main discrepancies are attributed to the simulation accuracy in replicating trigger requirements at low decay times. They are accounted for, both in the estimation of the relative fractions $f_i$ and of decay-time biases, by inflating uncertainties by a factor of $\sqrt{\chisqndf}$, separately for each decay-time bin. These discrepancies affect prompt and secondary decays similarly, and to a substantial extent cancel out in the computation of $f_i$ fractions.
\begin{figure}
    \centering
    \includegraphics[width=0.48\linewidth]{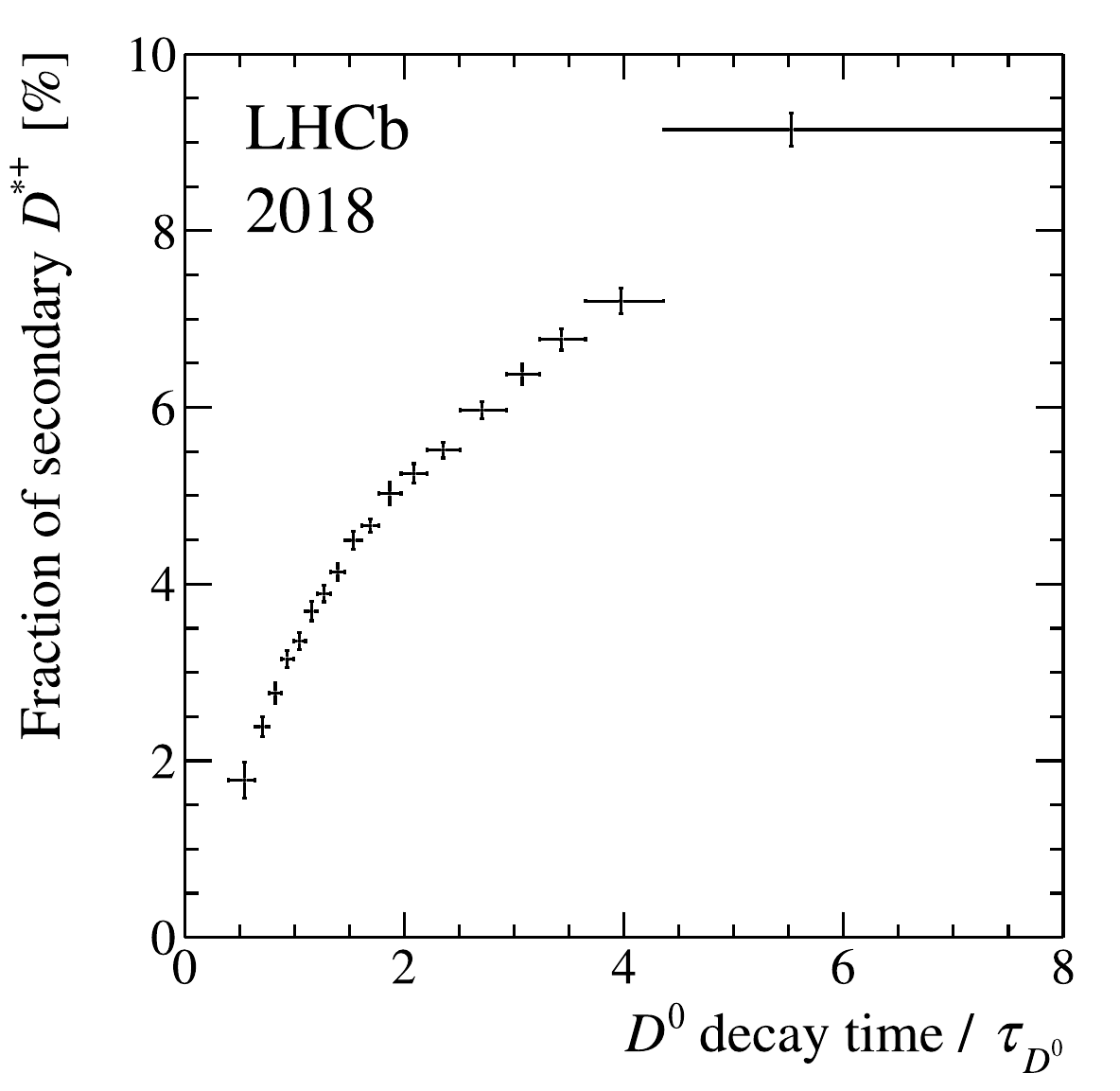}
    \includegraphics[width=0.48\linewidth]{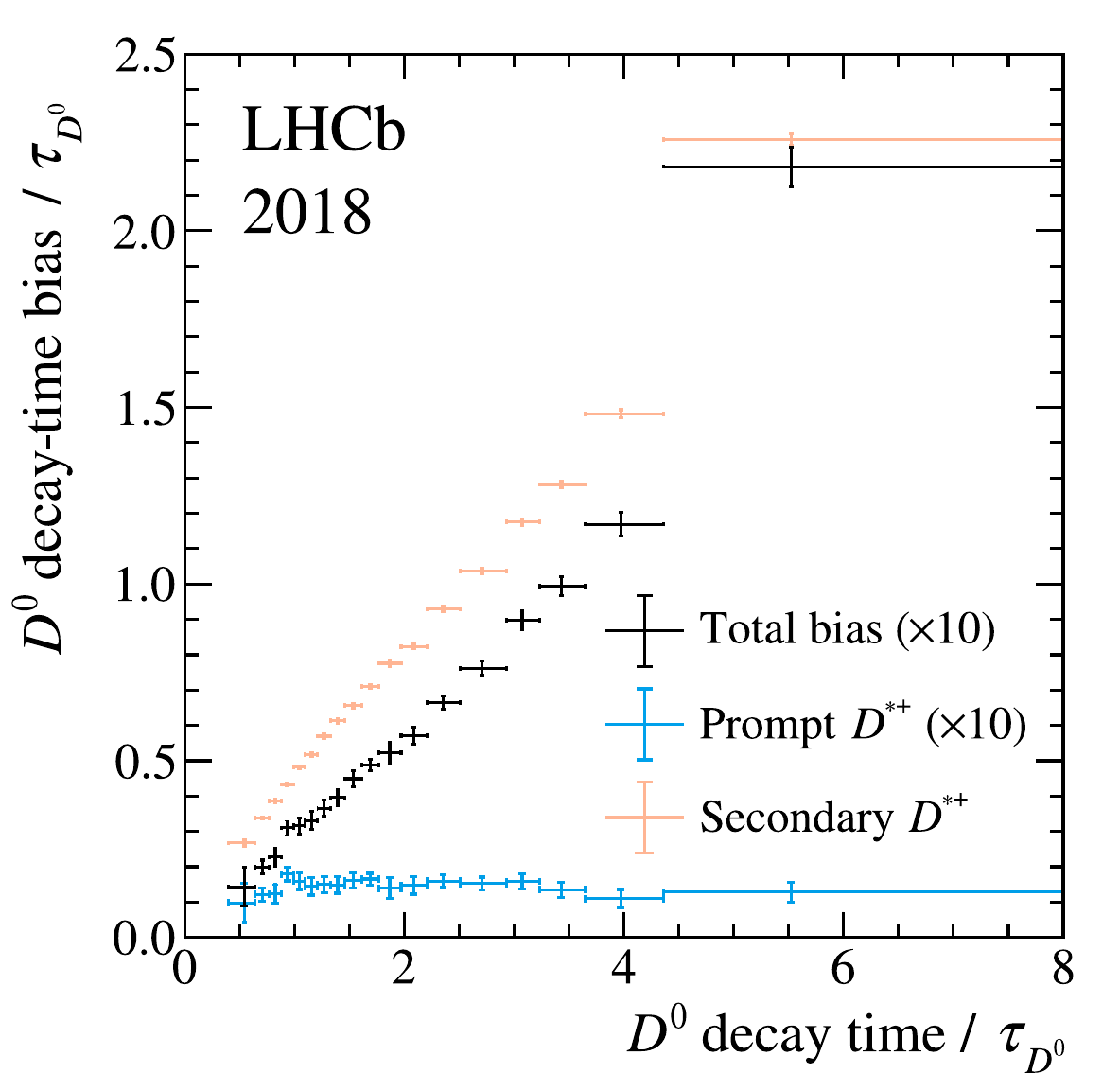}
    \caption{(Left) Relative fraction of \Dstarp candidates from $b$ hadrons as a function of the \Dz decay time. (Right) Total decay-time bias as a function of the \Dz decay time. Values of the decay-time bias for the prompt and secondary components are also shown. The \Dstarp candidates are required to have $\IP(\Dz)$ less than $60\mum$.}
    \label{fig:time_bias}
\end{figure}
The measured values of the total bias $\langle \delta t\rangle_i$, with the separate contributions from prompt and secondary decays, are shown in \cref{fig:time_bias}, together with the measured relative fractions, $f_i$, of secondary decays. The uncertainties on the $\langle \delta t\rangle_i$ parameters are also determined from the simulated samples of prompt and secondary decays, and include both statistical and systematic contributions. These uncertainties are correlated and the associated covariance matrix is used in the fit of the oscillation parameters, described in the next section.

\section{Determination of oscillation parameters}\label{sec:mix_fit}
The decay-time dependence of the WS-to-RS ratio is fitted to the observed ratios, simultaneously for the two \Dz final states, to determine the mixing and \CP-violation parameters. 
In the following, the uncertainties of measured quantities are indicated by $\sigma$.
The minimized $\chi^2$ is
\begin{equation}\label{eq:chi2}
	\chi^2=\sum_{j,y,f} \left( \frac{{r}_{jy}^{f} - R_{jy}^{f}}{\epsilon \, \sigma({r}_{jy}^{f})}\right)^2  + \chi^2_{\text{nuis}}\,,
\end{equation}
where the sum spans over all the 18 decay-time intervals, $j$, three data-taking periods, $y$, and the two \Dz final states, $f\in\{+,-\}$.
The terms ${r}_{jy}^{f}$ are the measured raw ratios, while their associated uncertainties are indicated with $\sigma({r}_{jy}^{f})$.
Small generic mismodeling found in the \Dstarp mass fits is accounted by an uncertainty inflation factor, $\epsilon = \sqrt{1.06}$, chosen such that the average reduced \chisq of the fits to the \Dstarp mass is equal to unity as explained in \cref{sec:mass_fit}.
The expected WS-to-RS ratios, $R^{\pm}_{jy}$, are written accounting for corrections as 
\begin{equation}
\begin{aligned}[b]
R^{\pm}_{jy} \equiv \big[& R_{_{\kaon\pion}} (1\pm A_{_{\kaon\pion}})+\sqrt{R_{_{\kaon\pion}} (1\pm A_{_{\kaon\pion}} )}\;(c_{_{\kaon\pion}}\pm\Delta c_{_{\kaon\pion}})\;\langle T \rangle_{jy}^\pm \\
&+(c_{_{\kaon\pion}}^{\prime}\pm \Delta c_{_{\kaon\pion}}^{\prime})\;\langle T^2\rangle_{jy}^{\pm}\big]\times \left(1\pm 2A_{jy}^\pm - C\right)+D,
\label{eq:Rpmty}
\end{aligned}
\end{equation}
where $R_{\kaon\pion}$, $c_{\kaon\pion}$, $c_{\kaon\pion}^{\prime}$, $A_{\kaon\pion}$, $\Delta c_{\kaon\pion}$ and $\Delta c_{\kaon\pion}^{\prime}$ are the six parameters of interest. 
The terms $\langle T^{(2)}\rangle_{jy}^{\pm}$ are the corrected average values of the (squared) decay time, defined as 
\begin{equation}
    \langle T^{(2)}\rangle _{jy}^\pm \equiv \left(\langle t^{(2)}\rangle_{jy}^\pm -   \langle \delta T^{(2)}\rangle_{jy}\right) \cdot S^{(2)},
\end{equation} 
where $\langle t^{(2)}\rangle_{jy}^\pm$ are the measured average (squared) decay times and the terms $\delta\langle T^{(2)}\rangle _{jy}$ are the nuisance fit parameters associated to the measured biases of the decay time.
The scale factor parameter $S$ accounts for the uncertainty of the $m_\Dz/\tau_\Dz$ ratio.
The $A_{jy}$ term in \cref{eq:Rpmty} is the instrumental asymmetry, defined as 
\begin{equation}\label{eq:Araw}
   A_{jy}^{\pm} \equiv A_{jy}^{\kaon\kaon} - A^{d}_{\kaon\kaon} - \delta y \cdot \langle T \rangle _{jy}^{\pm}\,,
\end{equation}
where $A_{jy}^{\kaon\kaon}$, $A^{d}_{\kaon\kaon}$ and $\delta y$ are the fit nuisance parameters associated to the measured raw asymmetry of the \decay{\Dz}{\Kp\Km} signal candidates (weighted to the RS sample), the \CP asymmetry in the decay \decay{\Dz}{K^+K^-} and the corresponding time-dependent \CP-violating asymmetry, respectively.
The nuisance parameter $C$ in \cref{eq:Rpmty} is the relative fraction of WS signal candidates discarded with the removal of the RS-WS common candidates, while $D$ is the nuisance parameter that accounts for the bias caused by the residual small contamination of doubly misidentified RS candidates.
Nuisance parameters are constrained to be within uncertainties of their measured values by the term
\begin{align}
	\chi^{2}_{\text{nuis}} =  &\sum_{y,i,j}(\delta \theta_{y}^{i}- \delta \Theta_{y}^i) [\text{Cov}^{-1}_y(\delta \theta)]^{ij}(\delta \theta_{y}^j- \delta \Theta_{y}^j) +  \left( \frac{s-S}{\sigma(s)}\right)^2  + \left( \frac{c-C}{\sigma(c)}\right)^2  \nonumber \\
&+ \left( \frac{d-D}{\sigma(d)}\right)^2 + \sum_{t,y}  \left( \frac{a^{\kaon\kaon}_{jy} -A^{\kaon\kaon}_{jy}}{\sigma(a^{\kaon\kaon}_{jy})}\right)^2 + \left( \frac{a^{d}_{\kaon\kaon} - A^{d}_{\kaon\kaon}}{\sigma(a^{d}_{\kaon\kaon})}\right)^2 + \left( \frac{\Delta Y - \delta y}{\sigma(\Delta Y)}\right)^2,\label{eq:nuisance}
\end{align}
where $\delta \theta_y$ and $\text{Cov}_y(\delta \theta)$ are the measured vectors of decay-time biases for different data-taking periods, defined as $\delta \theta_{y} \equiv [ \langle \delta t\rangle_{1y},  \langle \delta t^2\rangle_{1y},  \langle \delta t\rangle_{2y},  \langle \delta t^2\rangle_{2y}, \cdots]$, and their covariance matrices, respectively, while $\delta \Theta$ is the vector of nuisance fit parameters associated to the determined  bias values of the decay time, similarly defined as
\mbox{$\delta \Theta_{y} \equiv [ \langle \delta T\rangle_{1y},  \langle \delta T^2\rangle_{1y},  \langle \delta T\rangle_{2y},  \langle \delta T^2\rangle_{2y}, \cdots]$}.
The term $a^{\kaon\kaon}_{jy}$ is the measured raw asymmetry in the \decay{\Dz}{\Kp\Km} signal candidates, $c$ is the measured fraction of WS signal candidates discarded with the removal of the RS-WS common candidates, and $d$ is the measured bias caused by the residual contamination from doubly misidentified RS candidates. 
The values and associated uncertainties of the external inputs $a^{d}_{\kaon\kaon}$, $\Delta Y$, $m_\Dz$, $\tau_\Dz$ are taken from Refs.~\cite{HFLAV21, PDG2022}:
$\adkk = (4.5 \pm 5.3) \times 10^{-4}$, $\Delta Y = (-0.89 \pm 1.13) \times 10^{-4}$, $m_{_{\Dz}} = 1864.84 \pm 0.05 \mevcc$ and $\tau_{_{\Dz}} = 410.3 \pm 1.0 \fs$.
Finally, the value of $s$ is 1.0 and its uncertainty is determined as $\sigma^2(s)= \sigma^2(m_{_{\Dz}})/m^2_{_{\Dz}} + \sigma^2(\tau_{_{\Dz}})/\tau^2_{_{\Dz}}$.

The world average values of the external inputs \adkk and $\Delta Y$ are largely dominated by the \lhcb Run~2 measurements of the time-integrated and the time-dependent \CP violation in \decay{\Dz}{\Kp\Km} decays~\cite{LHCb-PAPER-2022-024,LHCb-PAPER-2020-045}, which mostly share the same candidates used for correcting the nuisance charge asymmetries in this measurement. However, the statistical correlation of mixing and \CP-violation observables with \adkk and $\Delta Y$ is small and is neglected in \cref{eq:chi2}. 
The statistical uncertainty of the \adkk measurement is dominated by the size of the \decay{\Dp}{\KS\pip} and \decay{\Ds}{\KS\Kp} calibration samples, used to precisely remove production and detection asymmetries\cite{LHCb-PAPER-2022-024}, while the measurement of $\AKpi$ reported in this article is largely dominated by that of the WS data sample. This results in a correlation of about 1\% between \AKpi and the two individual measurements of the time-integrated \CP violation in the \decay{\Dz}{\Kp\Km} decays of Ref.~\cite{LHCb-PAPER-2022-024}. The correlation of \DcKpi and \DcKpipr with the measurement of \DY~\cite{LHCb-PAPER-2020-045} is about 3\%, and its impact is completely negligible as also shown in \cref{tab:systematics}. 
For similar reasons a correlation of about 30\% between the \lhcb measurements of \adkk and $\Delta Y$ observables is also neglected in the minimization of the $\chi^2$ reported in \cref{eq:chi2}. 
An alternative fit, where the observables are slightly modified to be independent of these external inputs, such that the results can be directly used in global fits to the charm mixing and \CP-violation parameters~\cite{HFLAV21,LHCb-CONF-2022-003}, is described in \cref{app:no-constraints}. 
This alternative configuration also has the advantage that it can be used, under the assumption of negligible \CP violation in doubly Cabibbo-suppressed decays, to measure \adkk with improved precision. 

To search for undetected systematic uncertainties, the analysis is repeated on statistically independent data subsets chosen to be sensitive to specific sources of bias. These criteria include the data-taking year (2015–2016, 2017, 2018), the magnetic field orientation, the number of primary vertices in the event, the trigger category, the soft-pion momentum, the kinematic region of the soft pions (inside and on the border of the geometrical acceptance), and the output of the multivariate classifier used to remove ghost soft pions. The resulting variations of the measured parameters are consistent with statistical fluctuations, with $p$-values in the 9\%–86\% range. The stability of the results over the data-taking periods, trigger categories, and detector occupancy confirms the robustness of the analysis methodologies against any change due to different running and data-acquisition conditions, and aging of the \lhcb detector over the years. The compatibility of the results obtained by repeating the whole measurement for candidates collected with the two different magnet polarities probes the robustness and reliability of the corrections for the instrumental asymmetries. The consistency of the results in these two samples is considered a powerful validation of the analysis method, since without the application of the corrections a significant inconsistency is seen. The stability of the results as a function of the soft-pion momentum, and therefore as a function of the \Dz momentum, probes 
any subtle unknown effect due to track reconstruction algorithms, as well as any impact of the residual contamination from doubly misidentified RS candidates. Finally, repeating the measurement in different kinematic regions of the soft pions and in different intervals of the output of the multivariate classifier used to remove ghost soft-pion candidates is of paramount importance to ensure that removal of the ghost candidates is accurate and within the assessed uncertainties.

\section{Results and systematic uncertainties}\label{sec:results}
\begin{figure}
    \centering
    \includegraphics[width=0.65\linewidth]{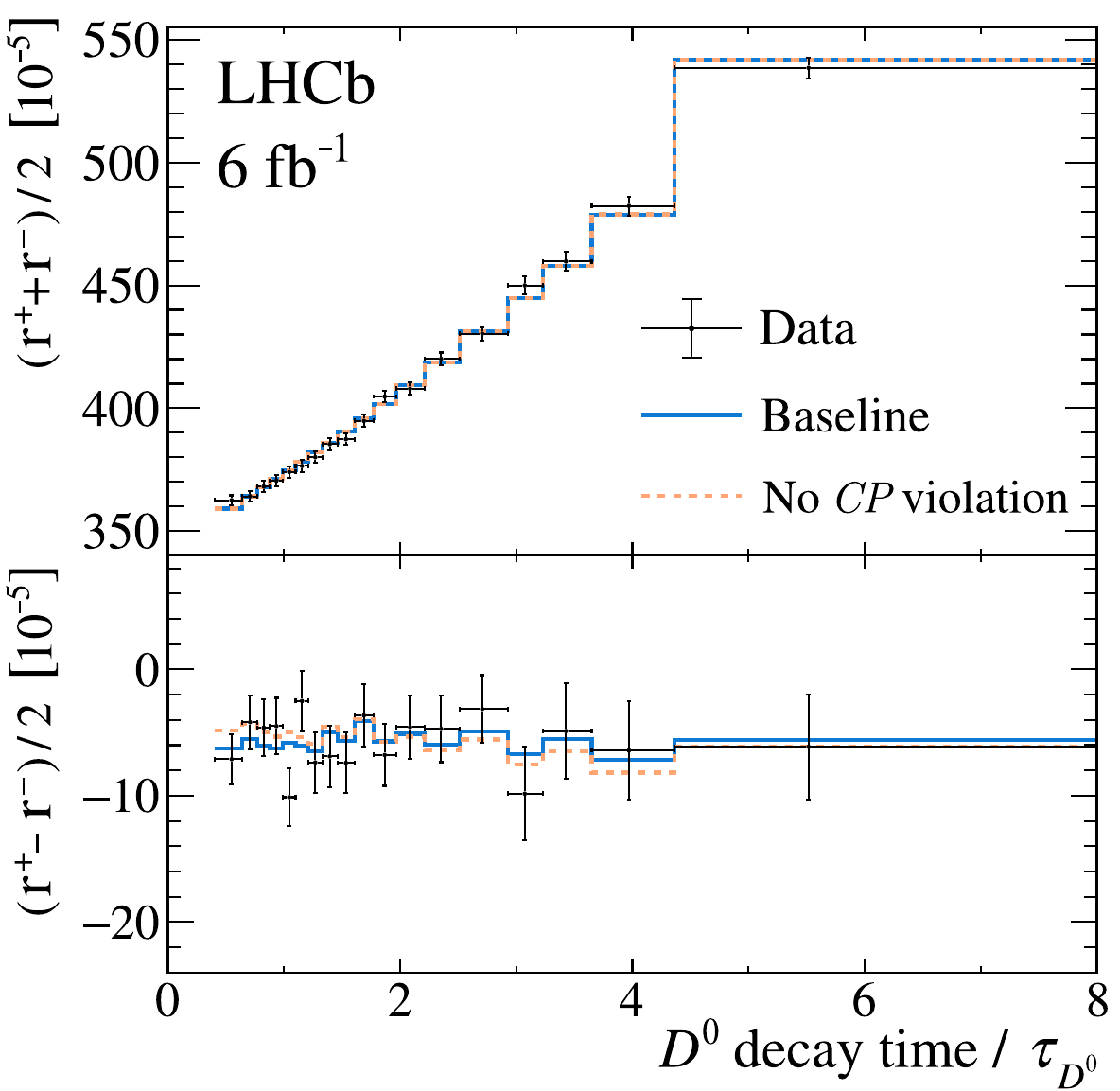}
    \caption{Half sum and half difference of measured WS-to-RS yields ratio for the \Kp\pim and \Km\pip final states as a function of decay time.   Projections of fits where \CP-violation effects are allowed (solid line) or forbidden (dotted line) are overlaid. The abscissa of each data point corresponds to the average decay time over the bin, the horizontal error bars delimit the bin, and the vertical error bars indicate the statistical uncertainties.}
    \label{fig:results}
\end{figure}

The fit results are presented in \cref{tab:results_run2}, where the uncertainties include both statistical and systematic contributions.
An ensemble of pseudoexperiments confirms that the extracted parameters are without perceivable bias and that the size of the parameter correlations are within expectations.
Fit projections, averaged over the three data-taking periods, are reported in \cref{fig:results}.
The obtained $p$-value of the fit is 0.91 (0.84), when the inflation factor $\epsilon$ is considered (not considered).
The fit is repeated in the scenario where \CP violation is not allowed, by requiring $\AKpi = 0$, $\DcKpi = 0$ and $\DcKpipr=0$.
The consistency of the data with the hypothesis of \CP symmetry is determined from the change in \chisq between the fits assuming \CP conservation and allowing for \CP violation.
The results are compatible with the hypothesis of \CP symmetry with a $p$-value of $0.57$.
The significance of the quadratic term in the decay-time-dependent ratio is similarly evaluated by repeating the fit with and without fixing \cKpipr to zero.
The difference in the $\chi^2$ value gives a significance of 3.4 standard deviations against the hypothesis of $\cKpipr = 0$.
This is the first measurement to have significant sensitivity to the quadratic term.
%%%%%%%%%%%%%%%%%%%%%%%%%
\begin{table}[tb]
\centering
    \caption{Results of the fit to the time dependence of the WS-to-RS ratio. Uncertainties and correlations include both statistical and systematic contributions.}
    \label{tab:results_run2}
\begin{tabular}{LRRRRRRR}
    \toprule
    \multicolumn{2}{c}{Parameters} & \multicolumn{6}{c}{Correlations [\%]} \\
    &  & \RKpi & \cKpi & \cKpipr & \AKpi & \DcKpi & \DcKpipr\\
    \midrule
    \RKpi & (343.1 \pm 2.0)\times 10^{-5}& 100.0 & -92.4 & 80.0 & 0.9 & -0.8 & 0.1\\
    \cKpi & (51.4 \pm 3.5)\times 10^{-4}&  & 100.0 & -94.1 & -1.4 & 1.4 & -0.7\\
    \cKpipr & (13.1 \pm 3.7)\times 10^{-6}&  &  & 100.0 & 0.7 & -0.7 & 0.1\\
    \AKpi & (-7.1 \pm 6.0)\times 10^{-3}&  &  &  & 100.0 & -91.5 & 79.4\\
    \DcKpi & (3.0 \pm 3.6)\times 10^{-4}&  &  &  &  & 100.0 & -94.1\\
    \DcKpipr & (-1.9 \pm 3.8)\times 10^{-6}&  &  &  &  &  & 100.0\\
    \bottomrule
\end{tabular}
\end{table}
%%%%%%%%%%%%%%%%%%%%
%%%%%%%%%%%%%%%%%%%%
The systematic uncertainties are summarized in \cref{tab:systematics}, where the contribution 
from each source is obtained by repeating the fit with the associated nuisance parameters fixed to their best-fit values and subtracting the resulting covariance matrix from the one of the unconstrained fit. 
The ``mass modeling'' contribution accounts for possible imperfections of the used empirical PDFs and is determined by repeating the fit by fixing the inflation factor $\epsilon$ to 1. 
The ``ghost soft pions'' contribution quantifies the impact on the analysis of the uncertainty on this component, and is determined by repeating all \mDp mass fits fixing the parameters of the PDF of the ghost component, in each decay-time interval, to those obtained in the baseline fits where they are free to float.
The ``instrumental asymmetry'' contribution refers to the uncertainties on the nuisance parameters 
$A_{jy}^{\kaon\kaon}$, which account for the statistical uncertainties in measuring the raw asymmetry in the \decay{\Dz}{\Kp\Km} signal candidates, and is relevant only for the \CP-violation parameters.
The contributions from ``$\adkk$ external input'' and ``$\Delta Y$ external input'' account for the uncertainties of the world-average values of these observables. The contributions ``doubly misidentified background'' and  ``removal of common [WS-RS] cand.'' refer to the uncertainty related to the estimate of the bias caused by the doubly misidentified RS candidates and to the uncertainty on the relative fraction of discarded WS candidates due to the removal of common WS-RS candidates. They are determined by repeating the fit with the nuisance parameters $C$ and $D$, respectively, fixed to their best-fit values. The item ``decay-time bias'' accounts for the uncertainties in the determination of the decay-time biases, which include biases from trigger-induced effects, uncertainties in the PV and DV resolutions, misalignments in the vertex detector, the composition of the simulated sample of secondary \Dstarp decays, the finite size of the simulation sample, and inflation of the uncertainty to account for local disagreements in the template fit. The systematic uncertainty is determined by repeating the fit while fixing the values of the $\langle\delta t \rangle_{jy}$ terms to their best-fit values.  The last contribution in the table refers to the scale factor $s$, which accounts for the uncertainty on the knowledge of the $m_{\Dz}/\tauDz$ ratio. 
The leading systematic uncertainties are those associated with the \Dstarp mass fit and the modeling of the ghost background; the results are dominated by statistical uncertainties.
\begin{table}
\centering
    \caption{Summary of the statistical and systematic uncertainties.
            A dash is used to indicate values below $0.1$ in the relevant units for that column.}
    \label{tab:systematics}
\begin{tabular}{lcccccc}
\toprule
\multirow{2}{*}{Source} & $\RKpi$ & $\cKpi$ & $\cKpipr$ & $\AKpi$ & $\DcKpi$ & $\DcKpipr$\\
 & [$10^{-5}$] & [$10^{-4}$] & [$10^{-6}$] & [$10^{-3}$] & [$10^{-4}$] & [$10^{-6}$]\\
\midrule
mass modeling & 0.5 & 0.8 & 0.9 & 1.4 & 0.8 & 0.8\\
ghost soft pions & 0.4 & 0.8 & 0.8 & 1.1 & 0.8 & 1.1\\
instrumental asymmetry & -- & -- & -- & 1.2 & 0.7 & 0.7\\
$\adkk$ external input & -- & -- & -- & 1.1 & -- & --\\
$\Delta Y$ external input & -- & -- & -- & -- & 0.1 & 0.1\\
doubly misidentified background & 0.1 & 0.1 & 0.1 & -- & -- & --\\
removal of common [WS-RS] cand. & 0.2 & -- & -- & -- & -- & --\\
decay-time bias & 0.1 & 0.2 & 0.1 & 0.1 & -- & -- \\
$m_{\Dz}/\tau_{\Dz}$ external input & -- & 0.1 & 0.1 & -- & -- & --\\
\midrule
total systematic uncertainty & 0.7 & 1.1 & 1.2 & 2.4 & 1.3 & 1.4\\
statistical uncertainty & 1.9 & 3.3 & 3.5 & 5.5 & 3.3 & 3.5\\
\midrule
total uncertainty & 2.0 & 3.5 & 3.7 & 6.0 & 3.6 & 3.8\\
\bottomrule
\end{tabular}
\end{table}

The results presented in this article extend and supersede the measurement of mixing and \CP-violation parameters with promptly produced \decay{\Dz}{\Kp\pim} decays published in Ref.~\cite{LHCb-PAPER-2017-046}  to the full \lhc Run~2 data sample, by adding data collected during 2017 and 2018, and by reanalyzing data collected in 2015 and 2016 with improved methodologies. The analysis of Ref.~\cite{LHCb-PAPER-2017-046} used data collected during \lhc Run~1 (2011--2012) combined with those from 2015 and 2016.
Since the 2015 and 2016 samples are used in both this analysis and the previous one, and separate results for Run~1 and~2 data were not presented in Ref.~\cite{LHCb-PAPER-2017-046}, the measurements cannot be combined naively.
Furthermore, the results reported in Ref.~\cite{LHCb-PAPER-2013-053}, based on Run~1 data alone, cannot be used, since the bias due to the contamination of ghost soft pions was not accounted for at that time.
%%%
To obtain results using LHCb's full available sample of prompt \decay{\Dz}{\Kp\pim} decays, the measurement of the full LHC Run~2 data sample, presented in this article, is combined with the previous \lhcb measurement based on the data collected during 2011 and 2012~\cite{LHCb-PAPER-2017-046}, by minimizing a total  $\chi^2 = \chi^2_{\textrm{Run1}} + \chi^2_{\textrm{Run2}}$, where the term $\chi^2_{\textrm{Run2}}$ is that described in \cref{eq:chi2}, while the term $\chi^2_{\textrm{Run1}}$ is recomputed using internal \lhcb documentation, and exactly matches the analysis of Ref.~\cite{LHCb-PAPER-2017-046}, including all systematic uncertainties and nuisance parameters.
There are no correlations between the two measurements and the two $\chi^2$ terms are treated as independent.
The final \lhcb Run~1 and Run~2 results are consistent with each other, and the combination reported in \cref{tab:results_run1+2} improves the final uncertainties compared to Run~2 alone by about 7\%. 
\begin{table}
\centering
    \caption{Results of the combination of the measurements using Run~1 and Run~2 data. Uncertainties and correlations include both statistical and systematic contributions.
             }
    \label{tab:results_run1+2}
\begin{tabular}{LRRRRRRR}
\toprule
\multicolumn{2}{c}{Parameters} & \multicolumn{6}{c}{Correlations [\%]} \\
&  & $\RKpi$ & $\cKpi$ & $\cKpipr$ & $\AKpi$ & $\DcKpi$ & $\DcKpipr$\\
\midrule
\RKpi & (342.7 \pm 1.9)\times 10^{-5} & 100.0 & -92.7 & 80.3 & 0.9 & -0.7 & 0.2\\
\cKpi & (52.8 \pm 3.3)\times 10^{-4} &  & 100.0 & -94.2 & -1.3 & 1.2 & -0.7\\
\cKpipr & (12.0 \pm 3.5)\times 10^{-6} &  &  & 100.0 & 0.7 & -0.7 & 0.2\\
\AKpi & (-6.6 \pm 5.7)\times 10^{-3} &  &  &  & 100.0 & -91.9 & 79.7\\
\DcKpi & (2.0 \pm 3.4)\times 10^{-4} &  &  &  &  & 100.0 & -94.1\\
\DcKpipr & (-0.7 \pm 3.6)\times 10^{-6} &  &  &  &  &  & 100.0\\
\bottomrule
\end{tabular}
\end{table}
%%%%%%%%%%%%
\section{Conclusions}
A measurement of the time-dependent ratio of the \decay{\Dz}{\Kp\pim} to \decay{\Dzb}{\Kp\pim} decay rates is performed at the \lhcb experiment using proton-proton collisions at a center-of-mass energy of $13\tev$, corresponding to an integrated luminosity of $6 \invfb$, collected from 2015 through 2018. The \Dz meson is required to originate from a prompt \decay{\theDstarp}{\Dz\pip} decay, such that its flavor at production is inferred from the charge of the soft pion. The measurement is performed simultaneously for the \Kp\pim and \Km\pip final states, allowing both mixing and \CP-violation parameters to be determined. 
Results are averaged with those obtained from Ref.\cite{LHCb-PAPER-2017-046} employing data recorded by the \lhcb experiment during 2011 and 2012 at a center-of-mass energy of $7 \tev$ and $8 \tev$, respectively, providing the legacy \lhcb Run~1 and Run~2 measurement. 

The results for all measured parameters are the most precise to date, and will significantly improve the precision of the world averages~\cite{HFLAV21,LHCb-PAPER-2021-033,LHCb-CONF-2022-003}.
The improvement factor of the total uncertainties with respect to the previous most precise set of measurements, also from LHCb~\cite{LHCb-PAPER-2017-046}, is in the range of 1.5--1.6 for both mixing and \CP-violation parameters, in line with the increase of the data sample size due to the inclusion of 2017 and 2018 data.  The most dominant source of systematic uncertainty of the previous measurement, due to the bias generated by the residual contamination of  \Dstarp decays originating from a $b$-hadron decay, is reduced by more than one order of magnitude. On average, the reduction of the total systematic uncertainties is about a factor of two, paving the way for even better precision in future measurements with larger data samples.
 
The parameter \AKpi provides a rigorous null test of the SM by probing \CP violation in the decay \decay{\Dz}{\Kp\pim}, and it is found to be consistent with \CP symmetry within an uncertainty of $5.7 \times 10^{-3}$.
The parameters \cKpi and \cKpipr constrain the values of the mixing parameters $x_{12}$ and $y_{12}$ and the phase \DKpi.
Since the value of $y_{12}$ is precisely known, particularly thanks to the recent measurement of the parameter $y_{\CP} = y_{12}\cos \phi^{\Gamma}_2$~\cite{LHCb-PAPER-2021-041} in singly Cabibbo-suppressed \decay{\Dz}{\Kp\Km} and \decay{\Dz}{\pip\pim} decays, the improvement in precision on \cKpi mainly impacts the accuracy in the determination of the strong phase \DKpi~\cite{LHCb-CONF-2024-004}, once the new results are combined with other relevant measurements from the charm and beauty sector following the approach of Ref.~\cite{LHCb-PAPER-2021-033}. 
The results reported in Ref.~\cite{LHCb-CONF-2024-004} point to a departure, exceeding 4 Gaussian-equivalent standard deviations, of \DKpi from the value of zero expected in the \suf symmetry limit, and they can provide insights on \suf breaking and rescattering at the energy scale of the charm quark mass. 
The precision of the new results also allows for the first time to have significant sensitivity to the quadratic term \cKpipr of the time-dependent expansion, which is found to deviate from zero with a significance of 3.4 Gaussian-equivalent standard deviations. 
Finally, the achieved accuracy on the parameter \DcKpi provides a clean test of \CP violation in the interference between \Dz mixing and decay, and will improve the knowledge of the dispersive mixing phase $\phi^M_2$ by about 16\%~\cite{LHCb-CONF-2024-004,charm-fitter}.

% Comment this in for paper drafts; do not include this in analysis note, conference and figure reports
\section*{Acknowledgements}
%
% These Acknowledgements valid from 3-May-2019
%
\noindent We express our gratitude to our colleagues in the CERN
accelerator departments for the excellent performance of the LHC. We
thank the technical and administrative staff at the LHCb
institutes.
We acknowledge support from CERN and from the national agencies:
CAPES, CNPq, FAPERJ and FINEP (Brazil); 
MOST and NSFC (China); 
CNRS/IN2P3 (France); 
BMBF, DFG and MPG (Germany); 
INFN (Italy); 
NWO (Netherlands); 
MNiSW and NCN (Poland); 
MCID/IFA (Romania); 
%MSHE (Russia); 
MICIU and AEI (Spain);
SNSF and SER (Switzerland); 
NASU (Ukraine); 
STFC (United Kingdom); 
DOE NP and NSF (USA).
We acknowledge the computing resources that are provided by CERN, IN2P3
(France), KIT and DESY (Germany), INFN (Italy), SURF (Netherlands),
PIC (Spain), GridPP (United Kingdom), 
%RRCKI and Yandex LLC (Russia), 
CSCS (Switzerland), IFIN-HH (Romania), CBPF (Brazil),
and Polish WLCG (Poland).
We are indebted to the communities behind the multiple open-source
software packages on which we depend.
Individual groups or members have received support from
ARC and ARDC (Australia);
Key Research Program of Frontier Sciences of CAS, CAS PIFI, CAS CCEPP, 
Fundamental Research Funds for the Central Universities, 
and Sci. \& Tech. Program of Guangzhou (China);
Minciencias (Colombia);
EPLANET, Marie Sk\l{}odowska-Curie Actions, ERC and NextGenerationEU (European Union);
A*MIDEX, ANR, IPhU and Labex P2IO, and R\'{e}gion Auvergne-Rh\^{o}ne-Alpes (France);
%RFBR, RSF and Yandex LLC (Russia);
AvH Foundation (Germany);
ICSC (Italy); 
%GVA, XuntaGal, GENCAT, Inditex, InTalent and Prog.~Atracci\'on Talento, CM (Spain);
Severo Ochoa and Mar\'ia de Maeztu Units of Excellence, GVA, XuntaGal, GENCAT, InTalent-Inditex and Prog. ~Atracci\'on Talento CM (Spain);
SRC (Sweden);
the Leverhulme Trust, the Royal Society
 and UKRI (United Kingdom).

\clearpage
\section*{Appendices}
\appendix
\section{Alternative parametrisations of the observables}
\label{app:alternative_param}

The parametrization of \cref{eq:ratio-parametrisation} is trivially related to that in Eq.~(73) of Ref.~\cite{Kagan:2020vri},
\begin{equation}
\label{eq:ratio-ks}
    \RKpipm(t) \approx \RKpipm
        + \sqrt{\RKpipm} \,\cWSpm\,t
        + \cWSprpm\, t^2,
\end{equation}
as follows,
\begin{equation}
    \begin{aligned}
        \RKpi &= \frac{\RKpip + \RKpim}{2},\\
        \AKpi &= \frac{\RKpip - \RKpim}{\RKpip + \RKpim},\\
        \cKpi  &= \frac{\cWSp + \cWSm}{2}, \\
        \DcKpi &= \frac{\cWSp - \cWSm}{2}, \\
        \cKpipr  &= \frac{\cWSprp + \cWSprm}{2},\\
        \DcKpipr &= \frac{\cWSprp - \cWSprm}{2}.
    \end{aligned}
\end{equation}
The parameters \cWSpm and \cWSprpm are often denoted as \yprimepm and $[\xprimepmsq + \yprimepmsq]/4$, respectively.
The definition of the phase \DKpi in \cref{eq:phase-definition} follows the convention adopted in Ref.~\cite{Kagan:2020vri}, and is related to those employed in Refs.~\cite{LHCb-PAPER-2021-033,LHCb-CONF-2022-003} and \cite{HFLAV21} as $\DKpi = \pi - \delta^{K\pi}_D = -\dKpi$.

In the phenomenological parametrization of \CP violation, defining the two mass eigenstates of the neutral charm meson system as $\ket{D_{1,2}} \equiv p \ket{\Dz} \pm q\ket{\Dzb}$, with $p$ and $q$ complex numbers satisfying $\lvert p \rvert^2 + \lvert q \rvert^2 = 1$ (\CPT invariance is assumed), and adopting the conventions that $\CP\ket{\Dz} = - \ket{\Dzb}$ and that $\ket{D_2}$ is the \CP-even eigenstate in the limit of \CP symmetry in charm mixing, the mixing parameters are defined as $x \equiv (m_2 - m_1) / \Gamma$ and $y \equiv (\Gamma_2 - \Gamma_1) / 2 \Gamma$, and are equal to $x_{12}$ and $y_{12}$ up to second order in $\sin(\phi^M_2 - \phi^\Gamma_2)$.
Finally, adopting the following convention for the weak phase responsible for \CP violation in the mixing~\cite{Kagan:2020vri},
\begin{equation}
    \begin{aligned}
        \phiKpi + \DKpi \equiv \arg\left(\frac{q \abf}{p \af}\right),\\
        \phiKpi - \DKpi \equiv \arg\left(\frac{q \abfb}{p \afb}\right),
    \end{aligned}
\end{equation}
the coefficients in \cref{eq:ratio-ks} can be expressed as
\begin{align}
    \cWSpm &\approx \left\lvert \frac{q}{p} \right\rvert^{\pm 1} [y \cos(\Delta_f \mp \phiKpi) + x \sin(\Delta_f \mp \phiKpi) ],\\
    \cWSprpm &\approx \frac{1}{4} \left(x^2+y^2 \right)\left\lvert \frac{q}{p} \right\rvert^{\pm 2}.
\end{align}
Second-order corrections to the expressions above, which are negligible within the current experimental precision, can be found in Ref.~\cite{Pajero:2021jev}.
The phase \phiKpi differs from the phase $\phi_2$ defined in Sec.~IV.B of Ref.~\cite{Kagan:2020vri}, and denoted as $\phi$ in Refs.~\cite{HFLAV21,LHCb-PAPER-2021-033,LHCb-CONF-2022-003}, by $O(10^{-6}) \rad$, \cf Sec. IV.C.2 of Ref.~\cite{Kagan:2020vri}.

\section{Results without external constraints}
\label{app:no-constraints}
The time-dependent fit, described in \cref{sec:mix_fit}, is repeated removing the nuisance parameters $A^{d}_{\kaon\kaon}$ and $\delta y$.
The Gaussian constraints associated with the external measurement of \adkk and \DY are consequently removed from~\cref{eq:nuisance}.
\Cref{eq:Rpmty} is  modified as follows,
\begin{align}
\nonumber R^{\pm}_{jy} \equiv \big[& R_{_{\kaon\pion}} (1\pm \tAKpi)+\sqrt{R_{_{\kaon\pion}} (1\pm \tAKpi )}\;(c_{_{\kaon\pion}}\pm \tDcKpi)\;\langle T \rangle_{jy}^\pm \\
&+(c_{_{\kaon\pion}}^{\prime}\pm \tDcKpipr)\;\langle T^2\rangle_{jy}^{\pm}\big]\times \left(1\pm 2A_{jy}^{\kaon\kaon} - C\right)+D,
\label{eq:alternative-params}
\end{align}
where the \CP asymmetry of the \decay{\Dz}{\Kp\Km} decay is absorbed into a new set of \CP-violation observables, which are approximately equal to
\begin{align}
\tAKpi &\approx \AKpi - 2\,\adkk,\\
\tDcKpi &\approx \DcKpi - \cKpi \adkk - 2\sqrt{\RKpi}\,\DY,\label{eq:dc-tilde}\\
\tDcKpipr&\approx\DcKpipr - 2 \cKpipr \adkk - 2\sqrt{\RKpi}\,\cKpi \,\DY,\label{eq:dcp-tilde}
\end{align}
where terms equal to $2\sqrt{\RKpi}\cdot \cKpi \cdot \DY (\langle T^2\rangle_{jy}^{\pm} - (\langle T\rangle_{jy}^{\pm})^2 )$ and to $-2\cKpipr \DY \langle T^2\rangle_{jy}^{\pm}\langle T\rangle_{jy}^{\pm}$ are neglected in \cref{eq:alternative-params}.
For completeness, note that \cref{eq:dc-tilde,eq:dcp-tilde} include terms that are of the same order as those neglected in \crefrange{eq:c}{eq:Dcpr}.
Higher-order corrections to \crefrange{eq:c}{eq:Dcpr} can be found in Ref.~\cite{Pajero:2021jev}.
The measurement of \tAKpi does not rely on the assumption of \CP symmetry in Cabibbo-favored \decay{\Dz}{\Km\pip} decays, contrary to all previous measurements of this decay channel.
The expressions for \crefrange{eq:A}{eq:Dcpr}, modified to account for \CP violation in Cabibbo-favored decays, can be found in Ref.~\cite{Pajero:2021jev}.

The combination of the present Run~2 measurement with the previous one of Run~1 from Ref.~\cite{LHCb-PAPER-2017-046}, described in \cref{sec:results}, is also slightly modified to account for the different approach utilized to remove the detection charge asymmetries. The Run~1 measurement did not use any experimental information from \decay{\Dz}{\Kp\Km} decays.
Therefore, the removal of the constraints on the nuisance parameters $A^{d}_{\kaon\kaon}$ and $\delta y$ does not affect the associated $\chi^2_{\textrm{Run1}}$ term.
Thus,  the parameters \tAKpi, \tDcKpi, and \tDcKpipr are used to probe \CP violation in Run~2 data, while the parameters \AKpi, \DcKpi, and \DcKpipr are used in Run~1 data.
The parameters \RKpi, \cKpi, and \cKpipr are unchanged and shared between the two $\chi^2$ terms.
Finally, the constraint on the parameter $s\propto\mDz/\tauDz$ is also removed, and the parameter $s$ is fixed to unity in the time-dependent fit.
The impact of this constraint on the results is negligible, as the relative uncertainty on this parameter, 0.25\%~\cite{PDG2022}, is much smaller than that on the measurement of the other observables.

The results without external constraints on \adkk and \DY and using this parametrization are reported in \cref{tab:results_run1+2_altpar}. They can be combined with other relevant measurements from the charm and beauty sector following the approach of Refs.~\cite{HFLAV21,LHCb-PAPER-2021-033}, with and without any desired constraint~\cite{LHCb-CONF-2024-004}.
For instance, this measurement acquires sensitivity to the \adkk parameter if \CP symmetry is assumed in doubly Cabibbo-suppressed $\decay{\Dz}{\Kp\pim}$ decays, as expected in the SM and in many of its extensions~\cite{Bergmann:1999pm,Grossman:2006jg}, namely $\AKpi = 0$.
With this assumption, an improvement in precision of about 10\% is obtained in the global fits to charm measurements for the determination of the \adkk observable~\cite{LHCb-CONF-2024-004}, with a similar improvement on the corresponding observable for the \decay{\Dz}{\pip\pim} decay, \adpipi, where a nonzero value with a significance exceeding 3 standard deviations has been recently found~\cite{LHCb-PAPER-2022-024}.
\begin{table}[tb]
\centering
\caption{Results of the combination of the measurements using Run~1 and Run~2 data without external constraints on \adkk, \DY, and $s$. Uncertainties and correlations include both statistical and systematic contributions.
         }
\label{tab:results_run1+2_altpar}
\resizebox{\textwidth}{!}{%
\begin{tabular}{LRRRRRRRRRR}
\toprule
\multicolumn{2}{c}{Parameters} & \multicolumn{9}{c}{Correlations [\%]} \\
&  & \RKpi & \cKpi & \cKpipr & \tAKpi & \tDcKpi & \tDcKpipr & \AKpi & \DcKpi & \DcKpipr\\
\midrule
\RKpi & (342.7 \pm 1.9) \times 10^{-5} & 100 & -92.7 & 80.3 & 0.8 & -0.7 & 0.0 & 0.3 & -0.2 & 0.2\\
\cKpi & (52.8 \pm 3.3) \times 10^{-4} &  & 100 & -94.3 & -1.4 & 1.3 & -0.6 & -0.5 & 0.4 & -0.4\\
\cKpipr & (12.0 \pm 3.5) \times 10^{-6} &  &  & 100 & 0.7 & -0.6 & 0.0 & 0.3 & -0.3 & 0.3\\
\tAKpi & (-8.2 \pm 5.9) \times 10^{-3} &  &  &  & 100 & -93.4 & 81.0 & 0.0 & 0.0 & 0.0\\
\tDcKpi & (3.2 \pm 3.6) \times 10^{-4} &  &  &  &  & 100 & -94.3 & 0.0 & 0.0 & 0.0\\
\tDcKpipr & (-2.0 \pm 3.8) \times 10^{-6} &  &  &  &  &  & 100 & 0.0 & 0.0 & 0.0\\
\AKpi & (-0.9 \pm 2.0) \times 10^{-2} &  &  &  &  &  &  & 100 & -93.8 & 81.1\\
\DcKpi & (-0.1 \pm 1.0) \times 10^{-3} &  &  &  &  &  &  &  & 100 & -94.3\\
\DcKpipr & (4.6 \pm 9.8) \times 10^{-6} &  &  &  &  &  &  &  &  & 100\\
\bottomrule
\end{tabular}%
}
\end{table}

% This should be taken out in the final paper
%\clearpage
%\input{supplementary}

\mciteErrorOnUnknownfalse
\addcontentsline{toc}{section}{References}
%\setboolean{inbibliography}{true}
\bibliographystyle{LHCb}
\bibliography{biblio,standard,LHCb-PAPER,LHCb-CONF,LHCb-DP,LHCb-TDR}

\newpage
% LHCb collaboration author list
% Data extracted on January 22nd, 2025 at 12:05pm for paper reference LHCb-PAPER-2024-008
\centerline
{\large\bf LHCb collaboration}
\begin
{flushleft}
\small
R.~Aaij$^{36}$\lhcborcid{0000-0003-0533-1952},
A.S.W.~Abdelmotteleb$^{55}$\lhcborcid{0000-0001-7905-0542},
C.~Abellan~Beteta$^{49}$,
F.~Abudin{\'e}n$^{55}$\lhcborcid{0000-0002-6737-3528},
T.~Ackernley$^{59}$\lhcborcid{0000-0002-5951-3498},
A. A. ~Adefisoye$^{67}$\lhcborcid{0000-0003-2448-1550},
B.~Adeva$^{45}$\lhcborcid{0000-0001-9756-3712},
M.~Adinolfi$^{53}$\lhcborcid{0000-0002-1326-1264},
P.~Adlarson$^{79}$\lhcborcid{0000-0001-6280-3851},
C.~Agapopoulou$^{13}$\lhcborcid{0000-0002-2368-0147}
% ...

{\footnotesize \it

% ...
$^{13}$Universit{\'e} Paris-Saclay, CNRS/IN2P3, IJCLab, Orsay, France\\
% ...
$^{36}$Nikhef National Institute for Subatomic Physics, Amsterdam, Netherlands\\
% ...
$^{45}$Instituto Galego de F{\'\i}sica de Altas Enerx{\'\i}as (IGFAE), Universidade de Santiago de Compostela, Santiago de Compostela, Spain\\
% ...
$^{49}$Physik-Institut, Universit{\"a}t Z{\"u}rich, Z{\"u}rich, Switzerland\\
% ...
$^{53}$H.H. Wills Physics Laboratory, University of Bristol, Bristol, United Kingdom\\
% ...
$^{55}$Department of Physics, University of Warwick, Coventry, United Kingdom\\
% ...
$^{58}$School of Physics and Astronomy, University of Glasgow, Glasgow, United Kingdom\\
$^{59}$Oliver Lodge Laboratory, University of Liverpool, Liverpool, United Kingdom\\
% ...
$^{67}$Syracuse University, Syracuse, NY, United States\\
% ...
$^{79}$Department of Physics and Astronomy, Uppsala University, Uppsala, Sweden, associated to $^{58}$\\
% ...
}
\end{flushleft}

\end{document}